\documentclass[lettersize,journal]{IEEEtran}
\usepackage{amsmath,amsfonts}
\usepackage{algorithmic}
\usepackage{algorithm}
\usepackage{array}
\usepackage[caption=false,font=normalsize,labelfont=sf,textfont=sf]{subfig}
\usepackage{textcomp}
\usepackage{stfloats}
\usepackage{url}
\usepackage{verbatim}
\usepackage{graphicx}
\usepackage{cite}
\usepackage{diagbox}
\usepackage{cleveref}
\usepackage{multirow}
\usepackage[table]{xcolor}
\begin{document}

\title{A Cost-Effective Submodule with Fault-Blocking Capability for Modular Multilevel Converters}

\author{Yu Liu, Shengbo Zhang
\thanks{Yu Liu and Shengbo Zhang are with the College 
	of Urban Transportation and Logistics, Shenzhen Technology University,
	Shenzhen, China.}
\thanks{Manuscript received November 5, 2026. Corresponding author: Yu Liu (liuyu@sztu.edu.cn).}}

\markboth{Journal of \LaTeX\ Class Files,~Vol.~14, No.~8, November~2026}%
{Shell \MakeLowercase{\textit{et al.}}: A Sample Article Using IEEEtran.cls for IEEE Journals}


\maketitle

\begin{abstract}
Modular multilevel converter (MMC) has become an attractive technology for high-power electronics applications thanks to its high energy efficiency, high power rating, high modularity, flexible scalability, and superior output waveform quality. It has attracted considerable research attention, and numerous submodule (SM) variants have been developed for different objectives. Among them, the half-bridge SM (HBSM) is the most widely adopted thanks to its simplicity, but it cannot interrupt DC-side short-circuit currents.  Various SM topologies with fault-blocking capability have been proposed. These SM topologies contain additional switches and passive components compared to the HBSM, leading to higher capital and operating costs. Aiming at these issue, an SM topology capable to block DC-side fault-currents with a reduced component count is proposed in this paper. The proposed SM includes two variants, one employs only two fully-controlled switches. An operation and control scheme for the proposed SM are also developed and presented in the paper. The functionality of the proposed SM is validated by experiment, and the value of the SM is demonstrated through a test simulation for an high-voltage direct-current (HVDC) transmission system. 
\end{abstract}


\section{Introduction}
\IEEEPARstart{M}{odular} multilevel converter (MMC) is an important technology in high-voltage direct current (HVDC) applications \cite{debnath2015,paez2019b}. It can also be employed in various applications such as flexible alternating current transmission system (FACTS), static compensator (STATCOM), motor drives, medium-voltage DC grids, and low-voltage DC distribution networks \cite{perez2021,port}. Its major features include highly modularity and scalability, high energy efficiency and voltage rating, low harmonics, and the absence of DC link \cite{debnath2015}. Due to these advantages, it has attracted great interests from researchers, and extensive developments have been made focusing on different aspects such as circuit topology, control scheme, modeling method, protection, application, etc. \cite{perez2021,port}. 

The basic operation principle of an MMC is to switch multiple submodules (SMs) connecting in series, so that a multilevel voltage waveform is generated at the ports of the MMC. Regarding the circuit topology of SM in MMC, several alternatives have been proposed in the literatures. Among the existing topologies, the half-bridge submodule (HBSM) and the full-bridge submodule (FBSM) are most widely adopted. The HBSM is the first proposed SM topology for MMC \cite{lesnicar2003}. In such topology, each SM contains one capacitor and two switches. Depending the states of the switches, the capactior can either be inserted or bypassed, resulting the output voltage of the SM equals to the capacitor voltage or zero, respectively. The HBSM topology is simple and widely used, however, it lacks the ability to handling DC side short-circuit fault since switch S$_2$ can only cut off current in one direction. 

Short-circuit fault on HVDC transmission lines may cause severe system failure and thus needs to be carefully dealt with. To cut off fault current, circuit breakers can be employed on the AC- or DC-side of MMC. However, they normally take several cycles to handle the fault, and this is not fast enough \cite{debnath2015}. One way to gain fault-handling capability for MMC is adding an extra switch, a capacitor, and two diodes to the half-bridge to form the diode-clamp submodule (DCSM) as proposed in \cite{li2013a}. The extra switch is connected in anti-series with the two switches in the half-bridge and enables the SM to cut off fault currents in the other direction. Fault current can thus be eliminated without circuit breaker, therefore, MMC can immediately resume normal power transmission for non-permanent fault. Note that the two capacitors are connected in series in the DCSM, reducing thereby the equivalent capacitance of an SM.  

Further exiting SM topologies with fault-blocking capability are formed by adding extra components to the HBSM topology. For instance, with two extra switches, the FBSM can fully bypass the capacitor when all switches are turned-off \cite{thitichaiworakorn2014}. The Mixed-Commutation Submodule (MCSM) \cite{nami2015} interconnects an HBSM and an FBSM to acquire the fault-blocking capability, and more voltage levels can be generated by one SM. Other topologies like the Full-Bridge Commutation Submodule (FBCSM) \cite{nami2015}, the Clamp-Double Submodule (CDSM) \cite{marquardt2010}, the Cross-Connected Submodule (CCSM) and the Parallel-Connected Submodule (PCSM) \cite{nami2013,chaudhuri2008} combines two FBSMs in cross-connected or in parallel with additional components, the abilities of generating multiple voltage levels and fault-blocking can also be achieved. Although each SM mentioned above can output more voltage levels than the HBSM, the total number of switches per normalized output voltage and the total number of switches in the conduction path of each SM are higher than the HBSM. The fault-handling ability of these topologies lead to higher capital cost and higher power loss compared to the HBSM \cite{nami2015}. For the above mentioned SM, the number of voltage level $N_{\rm v}$, the number of switches $N_{\rm s}$, the ratio of $N_{\rm v}$ over $N_{\rm s}$, the number of semiconductors in the current path when the SM is bypassed and inserted, $N^{\rm by}_{\rm s}$ and $N^{\rm in}_{\rm s}$, respectively, are summarized and listed in Table \ref{SM_Comparison} for comparison.

\begin{table}[h]
	\caption{Comparison of the HBSM and the Exsiting SMs with Fault-handling Capability \cite{nami2015}.}
	\label{SM_Comparison}    
	\centering
	\begin{tabular}{p{1cm}<{\centering} | p{1cm}<{\centering} | p{1cm}<{\centering} | p{1cm}<{\centering} | p{1cm}<{\centering} | p{1cm}<{\centering} }
		\hline \hline
		Topology & $N_{\rm v}$ & $N_{\rm s}$ & $\frac{N_{\rm s}}{N_{\rm v}}$ & $N^{\rm by}_{\rm s}$ & $N^{\rm in}_{\rm s}$ \\ \hline \hline
		
		HBSM & 2 & 2 & 1 & 1 & 1  \\ \hline
		
		FBSM & 3 & 4 & $\frac{4}{3}$ & 2 & 2 \\ \hline
		
		DCSM & 2 & 3 & $1.5$ & 2 & 2 \\ \hline
		
		CDSM & 4 & 5 & 1.25 & 3 & 3  \\ \hline
		
		MCSM & 4 & 6 & 1.5 & 3 & 3 \\ \hline
		
		CCSM & 5 & 6 & 1.2 & 3 & 3 \\ \hline
		
		PCSM & 5 & 6 & 1.2 & 4 & 4 \\ \hline
		
		\hline \hline
	\end{tabular}
\end{table}

It is important to note that in MMC-HVDC systems a three-phase MMC include multiple arms, and each arm normally contains hundreds of SMs to generate nearly sinusoidal voltage waveform. Therefore, any additional components in one single SM will result in hundreds or even thousands of extra components in the entire MMC. This will drastically increase the building cost and operation loss for the systems. Among the components, the switches, which are commonly implemented by insulated-gate bipolar transistors (IGBTs), normally have the highest unit price. Therefore, it is of great significance to reduce the number of components as much as possible, especially the switches, to decrease the building and operation cost, while retaining the fault-handing capability of the SMs. Aiming at this target, a novel topology of SM with fault-riding ability is proposed in this paper. In each proposed SM, only two switches are needed, and only one switch are included in the conduction path, making the building cost and power loss of the proposed SM comparable to the HBSM. 

This paper is organized as follows: Section \ref{topology} proposes the topology and its possible applications in MMC. The operation modes of the proposed SM is presented in Section \ref{modes}. The operation and balancing procedure for MMC using the proposed SM is then explained in details in Section \ref{procedure}. Then, the experimental validation of the proposed topology is presented and analyzed in Section \ref{procedure}. Thereafter, the value of the proposed SM is demonstrated through a test simulation. Finally, the conclusions are drawn in Section \ref{conclusion}.

\section{Topology of the Proposed Submodule} \label{topology}
\begin{figure}[htb]
	\centering
	\includegraphics[width=9cm]{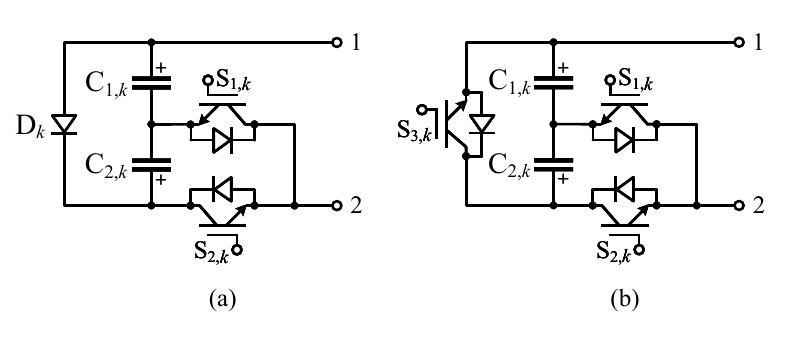}
	\caption{Circuit diagram of the proposed SM topologies: (a) SM variant 1; (b) SM variant 2.}
	\label{SM_Topology}
\end{figure}

The proposed SM is a passive circuit network with two variants as in Fig. \ref{SM_Topology}. The circuit diagram of the first variant is shown in Fig. \ref{SM_Topology}(a). For an arbitrary SM$_k$, where $k \in \mathbb{N}$, includes two ports marked as 1 and 2, two identical capacitors marked as C$_{1,k}$ and C$_{2,k}$, a diode marked as D$_k$, and two fully controlled power electronic switches S$_{1,k}$ and S$_{2,k}$ such as IGBT with free-wheeling diode connected in parallel as presented in this paper. port 1 is connected to the positive end of capacitor C$_{1,k}$, and the negative end of capacitor C$_{1,k}$ is connected to the negative end of capacitor C$_{2,k}$. This joint end is connected to port 2 through switch S$_{1,k}$. The anode and the cathode of diode D$_k$ are connected to the positive ends of capacitor C$_{1,k}$ and capacitor C$_{2,k}$, respectively. The positive end of capacitor C$_{2,k}$ is further connected to port 2 through switch S$_{2,k}$.  

The circuit diagram of the second variant is shown in Fig. \ref{SM_Topology}(b). This variant is basically the same as the first variant except the diode D$_k$ is replaced by a fully-controlled switch S$_{3,k}$ such as an IGBT as shown in the figure. Due to the unilateral conductivity of diode, current can only flow through the branch with the diode from port 1 to port 2 in the first variant. In the second variant, however, current can flow in both directions between the two ports.

\section{Operation Modes of the Proposed Submodule} \label{modes}
The proposed SMs have three operation modes. In mode 1, switch C$_{1,k}$ is closed and switch S$_{2,k}$ is open, capacitor C$_{1,k}$ is inserted to the external circuit, and the voltage across capacitor C$_{1,k}$ is applied to the ports of the SM. This mode is comparable to the insertion mode of the  SMs used in conventional MMC. In mode 2, switch C$_{1,k}$ is open and switch S$_{2,k}$ is closed, and the voltage at the ports of the SM is reduced near to zero. This mode is similar to the bypass mode of the conventional SMs. When both switches are turned off, the SMs work in Mode 3 in the blocked state. This mode can be adopted in the start-up stage or used to block fault current.

\subsection{Mode 1: Inserted State}
In this mode, two scenarios are considered. In the first scenario current $i_k$ flows from port 1 to port 2 and $i_k$ is defined as positive, while in the other scenario current $i_k$ flows from port 2 to port 1 and $i_k$ is defined as negative. 

\begin{figure}[htb]
	\centering
	\includegraphics[width=9cm]{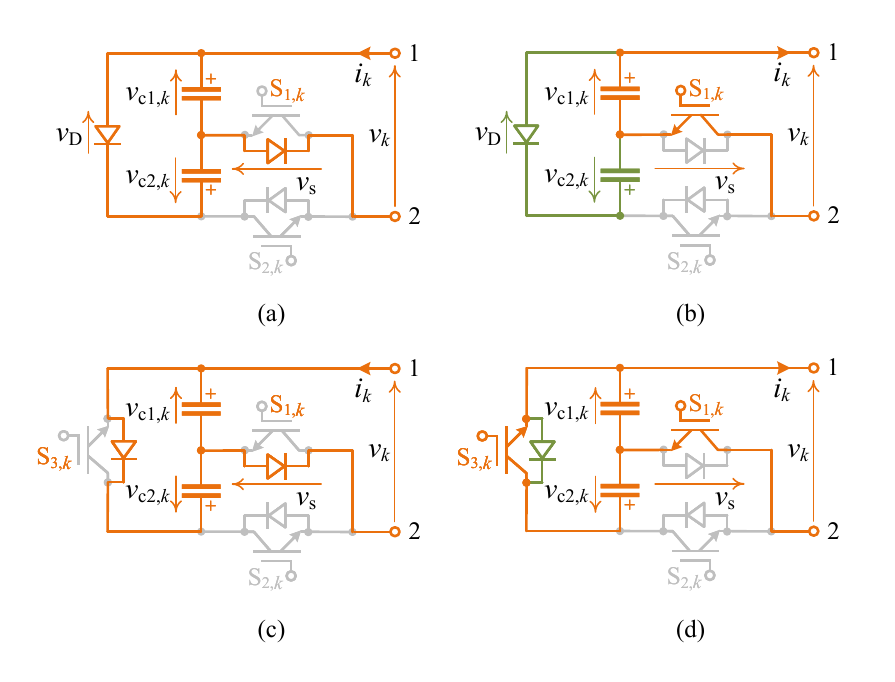}
	\caption{Equivalent circuit of the proposed SMs in Mode 1: (a) and (c) current flows from port 1 to 2; (b) and (d) current flows from port 2 to 1.}
	\label{Mode1}
\end{figure}

\paragraph{$i_k > 0$}
In this scenario, when the voltage across capacitor C$_{1,k}$, denoted as $v_{{\rm c1},k}$, is greater than the sum of the voltage across capacitor C$_{2,k}$ and the forward voltage drop across diode D$_k$, denoted as $v_{{\rm c2},k}$ and $v_{\rm D}$, respectively, then the power diode D$_k$ also conducts. Consequently, the current $i_k$ splits into two paths flowing through capacitors C$_{1,k}$ and C$_{2,k}$ respectively, and subsequently through switch S$_{1,k}$ to port 2. In this scenario, capacitors C$_{1,k}$ and C$_{2,k}$ are connected in parallel and being charged, and both $v_{{\rm c1},k}$ and $v_{{\rm c2},k}$ increase. The equivalent circuit of both SMs working in this scenario are represented by the orange part as shown in Fig. \ref{Mode1}(a) and (c). The voltage between port 1 and port 2 of the SM, denoted as $v_k$, equals the sum of voltage $v_{{\rm c1},k}$ and $v_{\rm S}$, which is the forward voltage drop of each switch and is neglected in this paper since $v_{\rm S}$ is much smaller than $v_{{\rm c1},k}$ during operation, i.e., 
\begin{equation}
	\label{Vc1}
	v_k = v_{{\rm c1},k} + v_{\rm S} \approx v_{{\rm c1},k}.
\end{equation}
Voltage $v_{{\rm c1},k}$ equals the sum of voltages $v_{{\rm c2},k}$ and $v_{\rm D}$, which is approximately equal to $v_{{\rm c2},k}$ because $v_{\rm D}$ is much smaller than $v_{{\rm c2},k}$ during operation and is neglected in this paper, i.e., 
\begin{equation}
	v_{{\rm c1},k} = v_{{\rm c2},k} + v_{\rm D} \approx v_{{\rm c2},k}.
\end{equation}

On ther other hand, if $v_{{\rm c1},k} < v_{{\rm c2},k}$, power diode D$_k$ is reversely biased and current $i_k$ flows entirely through capacitor C$_{1,k}$. Therefore, $v_{{\rm c1},k}$ increases whereas $v_{{\rm c2},k}$ remains constant. In this case, $v_k$ equals the sum of $v_{{\rm c1},k}$ and $v_{\rm S}$ as previously and \cref{Vc1} is still valid. Once $v_{{\rm c1},k} \geq v_{{\rm c2},k}$, D$_k$ begins to conduct and the SMs work as previously, i.e., $v_k = v_{{\rm c1},k}$. An can be seen, $v_{{\rm c1},k}$ and $v_{{\rm c2},k}$ can be automatically balanced in this scenario. 

\paragraph{$i_k < 0$}
In this scenario, current $i_k$ flows entirely through capacitor C$_{1,k}$ since the path through capacitor C$_{2,k}$ is blocked by D$_k$. Therefore, $v_{{\rm c1},k}$ decreases and $v_{{\rm c2},k}$ remains constant. The equivalent circuit of both SMs are represented by the orange part as shown in Fig. \ref{Mode1}(b) and (d). If $v_{{\rm c1},k} \ge v_{{\rm c2},k}$, D$_k$ will conduct, and capacitor C$_{2,k}$ will be charged by capacitor C$_{1,k}$ until $v_{{\rm c1},k} = v_{{\rm c2},k}$. For the extended SM, if either $v_{{\rm c1},k}$ or $v_{{\rm c2},k}$ is lower than $v^k_{\rm dc}$, the corresponding capacitor can be charged by the DES/DESS connected to the SM. On the contrary, if $v_{{\rm c1},k}$ is higher than the predefined value or $v_{{\rm c2},k}$, $v_{{\rm c1},k}$ can be decreased by through discharging to the connected DESS. Again, $v_k$ is equal to the sum of $v_{{\rm c1},k}$ and $v_{\rm S}$, which is approximately equal to $v_{{\rm c1},k}$ in this scenario, and \cref{Vc1} is valid.

As discussed above, capacitor C$_{1,k}$ is inserted to the external circuit in mode 1 regardless the direction of the current. The output voltage of the SM $v_k$ is approximately equal to the voltage across capacitor C$_{1,k}$. If external short-circuit fault occurs when $i_k < 0$, the fault current can be turned off by S$_{1,k}$. Although S$_{1,k}$ lacks the ability to turn off fault current in the other direction, the rapidly increasing voltage $v_{{\rm c1},k}$ due to the fault current in return suppress itself since capacitor C$_{1,k}$ is inserted in the circuit. Therefore, the SM has the ability to block fault current in this mode. Furthermore, the proposed SMs can automatically balance the capacitor voltages except in the case where $i_k < 0$ and $v_{{\rm c1},k} < v_{{\rm c2},k}$. 

\subsection{Mode 2: Bypassed State}
\begin{figure}[htb]
	\includegraphics[width=9cm]{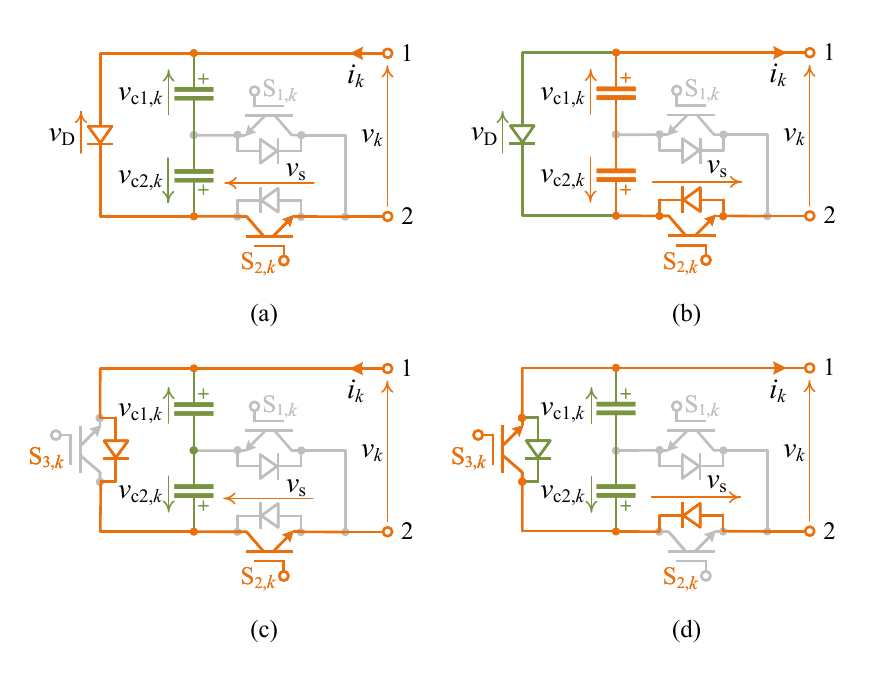}
	\caption{Equivalent circuit of the proposed SMs in Mode 2: (a) and (c) current flows from port 1 to 2; (b) and (d) current flows from port 2 to 1.}
	\label{Mode2}
\end{figure}

Similar to the previous mode, two scenarios regarding the current flow direction are considered in this mode. When current flows from port 1 to port 2 ($i_k \geq 0$) and $v_{{\rm c1},k} \geq v_{{\rm c2},k}$, the two capacitors are bypassed by the power diode. Current $i_k$ flows directly through the diode and switch S$_{2,k}$ to port 2, as shown in the orange part as shown in Fig. \ref{Mode2}(a) and (c). If $v_{{\rm c1},k} > v_{{\rm c2},k}$, additional current will flow from C$_{1,k}$ to C$_{2,k}$ until $v_{{\rm c1},k} = v_{{\rm c2},k}$, balancing the voltages of the two capacitors. The output voltage $v_k$ at the ports of the SM becomes zero when the voltage $v_{\rm S}$ across switch S$_{2,k}$  is neglected, i.e., $v_k = 0$. 

When $v_{{\rm c1},k} < v_{{\rm c2},k}$, $v_{\rm D}$ is blocked, the current flows through the branch consisting of the two capacitors. In this case, C$_{1,k}$ is charging and C$_{2,k}$ is discharging. Thus, $v_{{\rm c1},k}$ increases whereas $v_{{\rm c2},k}$ decreases, and the difference between $v_{{\rm c1},k}$ and $v_{{\rm c2},k}$ will be reduced until $v_{{\rm c1},k} = v_{{\rm c2},k}$. Thereafter, the power diode D$_k$ conducts and the capacitors are bypassed, the SMs work as in the previous case, and the capacitor voltages are balanced. 

When current flows from port 2 to port 1 ($i_k < 0$), it cannot pass the diode D$_k$. Instead, it flows through the branch including the two capacitors, as shown in the orange part in Fig. \ref{Mode2}(b) and (d). The capacitors are not essentially bypassed in this scenario. Instead, they are both inserted to the external circuit, C$_{1,k}$ is will be discharged and C$_{2,k}$ will be charged. Therefore, voltage $v_{{\rm c1},k}$ is offset by the voltage $v_{{\rm c2},k}$,  the output voltage $v_k$ of the SM equals the difference between the two capacitor voltages denoted as $\Delta v_{\rm c}$, i.e., 
\begin{equation}
	v_k = \Delta v_{\rm c} = v_{{\rm c1},k} - v_{{\rm c2},k}.
\end{equation}

During operation, the capacitor voltages are controlled close to each other, and this can be automatically achieved in most cases as discussed previously. As the capacitors are connected in anti-series, their voltages basically cancel each other out along the branch, resulting in $v_k \approx 0$ as if the capacitors were bypassed. Notice that, when $v_{{\rm c1},k} > v_{{\rm c2},k}$, the diode is conducting and the capacitor voltages will be automatically balanced. However, the reverse process cannot take place automatically since the path from C$_{2,k}$ back to C$_{1,k}$ is blocked by D$_k$. As $v_{{\rm c1},k}$ decreases and $v_{{\rm c2},k}$ increases, $v_k$ decreases below $v_{\rm S}$ or even 0 once $v_{{\rm c1},k} \leq v_{{\rm c2},k}$. Therefore, additional measurement needs to be taken to balance the capacitor voltages. For the extended SM, the voltage difference can be reduced by charging C$_{1,k}$ through the DES/DESS connected to ports 3 and 4.

In this mode, the capacitors in the proposed SM are bypassed when $i_k \geq 0$ and pseudo-bypassed when $i_k < 0$. In the previous case, external fault current can be turned-off by switch S$_{2,k}$. Although S$_{2,k}$ cannot switch off negative fault current, the current will rapidly increase $v_{{\rm c2},k}$, which in return will suppress the current itself. The pseudo-bypass state enables the SM to block external fault current because the capacitors are not actually bypassed. This is a major merit brought by the proposed SM. In the traditional half-bridge topology, fault current cannot be blocked when the SM is bypassed. To gain this blocking ability, at least one additional switch needs to be added to the half-bridge structure such as the diode-clamp SM presented in \cite{li2013}. Other existing SM topologies with fault blocking ability require more switches. In the proposed SM topology, however, only two switches are needed. This can help reduce the building cost and control complexity of MMC.

\subsection{Mode 3: Blocked State}
\begin{figure}[htb]
	\centering
	\includegraphics[width=9cm]{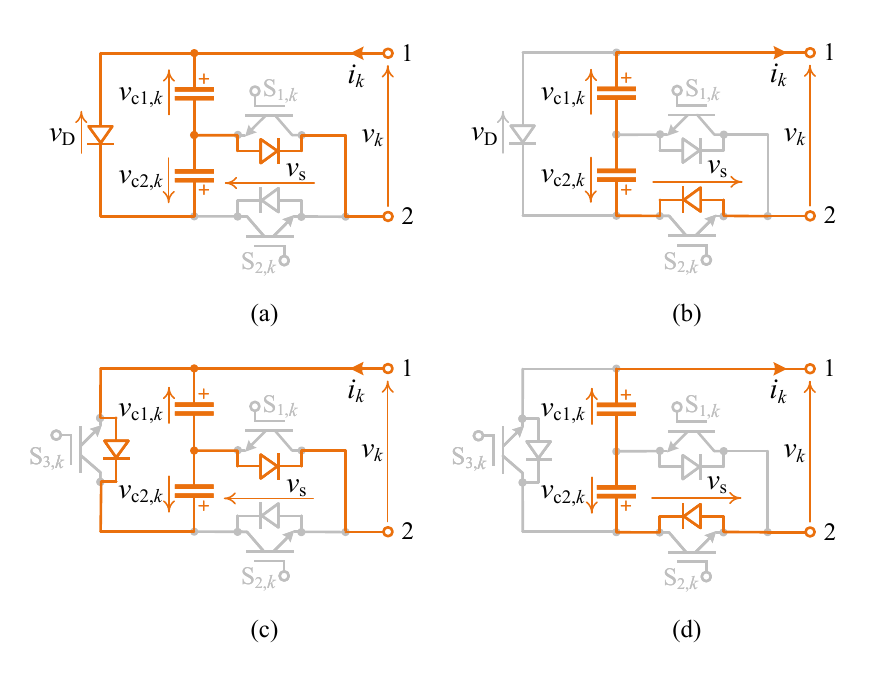}
	\caption{Equivalent circuit of the proposed SMs in Mode 3: (a) and (c) current flows from port 1 to 2; (b) and (d) current flows from port 2 to 1.}
	\label{Mode3}
\end{figure}

When both switches S$_{1,k}$ and S$_{2,k}$ are turned off, the SM operates in Mode 3 where the SMs are blocked. This mode can be adopted in the start up process or when short-circuit fault occurs. During the transient state, when current $i_k$ flows from port 1 to port 2 ($i_k \geq 0$) and $v_{{\rm c1},k} \geq v_{{\rm c2},k}$, D$_k$ conduct and the two capacitors are connected in parallel. The current flows through the two branches containing the capacitors, respectively, and then merges into one at the joint point of the capacitors. Thereafter, the current flows through the freewheeling diode of switch S$_{1,k}$ and then out of the SMs from port 2. Both capacitors are being charged by current $i_k$. Meanwhile, additional current flows from C$_{1,k}$ to C$_{2,k}$ until $v_{{\rm c1},k} = v_{{\rm c2},k}$. The output voltage of the SM $v_k$ is equal to $v_{{\rm c1},k}$ in this case. The equivalent circuit of the SMs are shown in the orange part in Fig. \ref{Mode3} (a) and (c). 

In this stage, both $v_{{\rm c1},k}$ and $v_{{\rm c2},k}$ increase. In the start-up process, $v_{{\rm c1},k}$ and $v_{{\rm c2},k}$ increase gradually in this mode. Once they reach the preset value, the SMs are ready for operation and this mode can be terminated. Under abnormal condition such as DC short circuit fault, the fault current will rapidly increase $v_{{\rm c1},k}$ and $v_{{\rm c2},k}$, which in return suppresses the fault current. As long as $v_{{\rm c1},k}$ and $v_{{\rm c2},k}$ equals the external voltage, $i_k$ is reduced to zero. On the contrary when $v_{{\rm c1},k} < v_{{\rm c2},k}$, D$_k$ is blocked and current $i_k$ flows entirely through capacitor C$_{1,k}$. Thus, $v_{{\rm c1},k}$ increases while $v_{{\rm c2},k}$ remains constant. Since capacitor C$_{1,k}$ is still connected to the two ports in parallel, $v_k = v_{{\rm c1},k}$. Once $v_{{\rm c1},k}$ reaches $v_{{\rm c2},k}$, D$_k$ begins to conduct and the SMs operate as in the previous scenario. 

If current flows from port 2 to port 1 ($i_k < 0$), it can only pass through the freewheeling diode of switch S$_{2,k}$ since both switches are turned off. The branch containing D$_k$ is also blocked for the current due to its unidirectional conductivity. Therefore, $i_k$ flows through the branch that consisted of the two capacitors connecting in series and then out of the SMs from port 1. Subsequently, $v_{{\rm c1},k}$ decreases and $v_{{\rm c2},k}$ increases. The equivalent circuit of the SMs in this scenario is shown by the orange part in Fig. \ref{Mode3} (b) and (d). If $v_{{\rm c1},k} > v_{{\rm c2},k}$, additional current flows from C$_{1,k}$ through D$_k$ to C$_{2,k}$, as shown in the green part in the figure, until $v_{{\rm c1},k} = v_{{\rm c2},k}$. Since the two capacitor are connected in anti-series, their voltages cancel each others out resulting $v_k = \Delta v_{\rm c}$. 

All operation modes and scenarios of the proposed SMs are summarized in Table \ref{Modes}. As can be seen from the table, the capacitor voltages can be automatically balanced in most scenarios during operation except three scenarios. In the first scenario, only capacitor C$_{1,k}$ is inserted and capacitor C$_{2,k}$ is bypassed. Therefore, the unbalanced capacitor voltages does not impact the output voltage $v_k$ of the SM. The voltage difference can be mitigated in other scenarios. The exceptional scenario that affiliated with Mode3, and the mode is used in start-up process or to block fault current under abnormal condition. Although the capacitor voltages cannot be automatically balanced, similar to the previous one, the unbalanced capacitor voltages does not impact the normal operation of MMC. In the exceptional scenario in Mode 2, however, both capacitors are inserted to the external circuit, and $v_{{\rm c1},k}$ is offset by $v_{{\rm c2},k}$. Since $v_k = \Delta v_{\rm c} = v_{{\rm c1},k} - v_{{\rm c2},k}$, $v_k$ becomes negative when $v_{{\rm c1},k} < v_{{\rm c2},k}$. In this case, the value of $|\Delta v_{\rm c}|$ needs to be reduced as much as possible to make $v_k$ around 0. Therefore, this exceptional scenario of Mode 2 is the only one for which additional measurement needs to be taken to balance the capacitor voltages.
\begin{table}[h]
	\caption{Summary of the operation modes and scenarios.}
	\label{Modes}    
	\centering
	\begin{tabular}{p{0.6cm}<{\centering} | p{0.6cm}<{\centering} | p{0.6cm}<{\centering} | p{0.6cm}<{\centering} | p{0.6cm}<{\centering} | p{1.1cm}<{\centering} | p{1.1cm}<{\centering}}
		\hline \hline
		Mode & S$_{1,k}$ & S$_{2,k}$ & $v_k =$ & $i_k$ & $v_{{\rm c1},k}$ & Automatic balance? \\ \hline
		
		\multirow{4}{*}{1} & \multirow{4}{*}{ON} & \multirow{4}{*}{OFF} & \multirow{4}{*}{$v_{{\rm c1},k}$} & \multirow{2}{*}{$\geq 0$} &  $ \geq v_{{\rm c2},k}$ & \multirow{3}{*}{Yes} \\ \cline{6-6}
		
		& & & & &  $ < v_{{\rm c2},k}$ &  \\ \cline{5-6}
		
		& & & & \multirow{2}{*}{$<0$} &  $ \geq v_{{\rm c2},k}$ &  \\ \cline{6-7}
		
		& & & & & $ < v_{{\rm c2},k}$ & No \\ 
		
		\hline
				
		\multirow{4}{*}{2} & \multirow{4}{*}{OFF} & \multirow{4}{*}{ON} & \multirow{4}{*}{$\Delta v_{\rm c}$} & \multirow{2}{*}{$\geq 0$} &  $\geq v_{{\rm c2},k}$ & \multirow{3}{*}{Yes} \\ \cline{6-6}
		
		& & & & & $ < v_{{\rm c2},k}$ &  \\ \cline{5-6}
		
		& & & & \multirow{2}{*}{$<0$} &  $ \geq v_{{\rm c2},k}$ &  \\ \cline{6-7}
		
		& & & & &  $ < v_{{\rm c2},k}$ & \textcolor{red}{No} \\
		
		\hline
		
		\multirow{4}{*}{2} & \multirow{4}{*}{OFF} & \multirow{4}{*}{OFF} & \multirow{2}{*}{$v_{{\rm c1},k}$} & \multirow{2}{*}{$\geq 0$} &  $\geq v_{{\rm c2},k}$ & \multirow{3}{*}{Yes} \\ \cline{6-6}
		
		& & & & & $ < v_{{\rm c2},k}$ &  \\ \cline{4-6}
		
		& & & \multirow{2}{*}{$\Delta v_{\rm c}$} & \multirow{2}{*}{$<0$} &  $ \geq v_{{\rm c2},k}$ &  \\ \cline{6-7}
		
		& & & & &  $ < v_{{\rm c2},k}$ & No \\
		\hline \hline
	\end{tabular}
\end{table}

For the extended SM connected to DES/DESS operating in this scenario, power can be drawn from the DES/DESS to charge capacitor C$_{1,k}$ to balance the capacitor voltages. For the proposed two-port SM operating in this scenario, the operation mode needs to be switched from Mode 2 to Mode 1, in which capacitor C$_{2,k}$ is bypassed and thus negative $v_k$ can be avoided. 

Low power loss comparable to HBSM since there is only one switch conducting in Mode 1 and Mode 2.

\section{Operation Procedure} \label{procedure}
\begin{figure}[htb]
	\includegraphics[width=9cm]{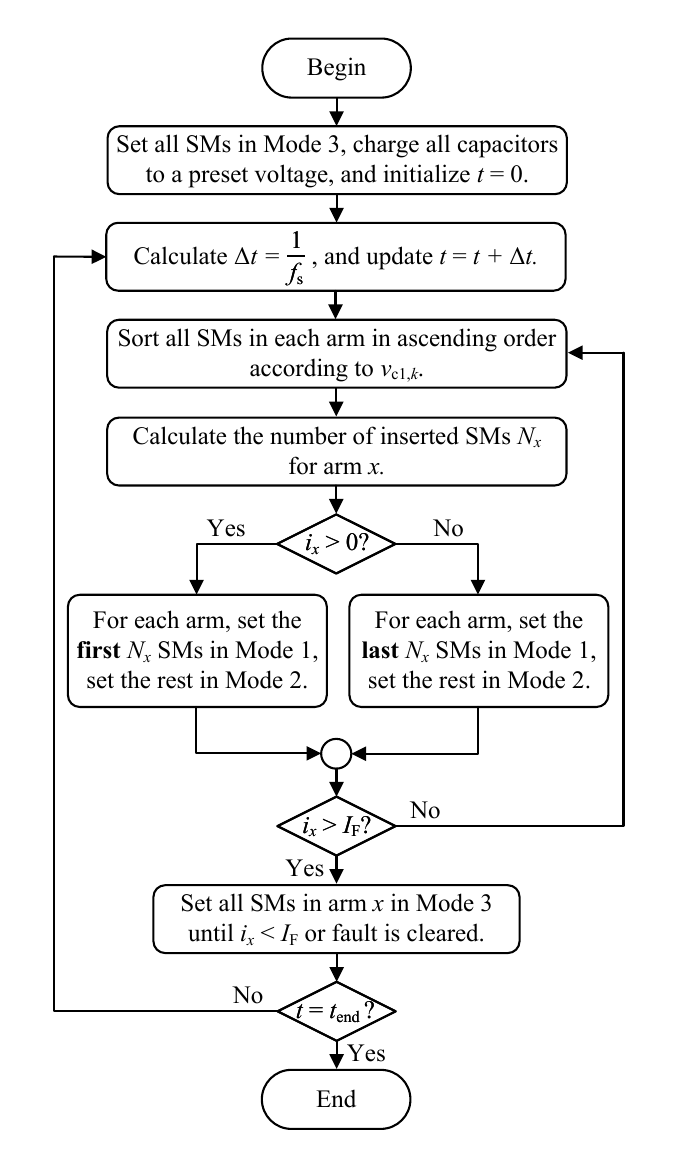}
	\caption{Operation procedure of the proposed MMC.}
	\label{Procedure}
\end{figure}

An operation procedure for the proposed SMs is designed and presented in this section. The procedure includes starting up the SMs, balancing the capacitor voltages, and contingency handling. The operation procedure is shown in the flow chart in Fig. \ref{Procedure}. The procedure repeats itself until the MMC is turned off.

\paragraph{Initializing SMs} Before putting the SMs into operation, they need to be properly started up. Above all, the capacitors in all SMs need to be charged to a preset voltage. To charge the capacitors, all SMs are set in Mode 3, and the SMs will be charged by the DC voltage source connected to the DC side of the MMC. If the SMs are not connected to DES or DESS, the initial voltages of all capacitors need to be set to the same value before charging so that they can reach the same final state. On the contrary, when the SMs are powered by DES or DESS, the final capacitor voltages can be independently regulated regardless of their initial states. At the start-up stage, the MMC needs to be disconnected to any AC source or utility grid at the AC side. After the SMs are charged to the preset voltage, they can be put into operation, and the moment $t$ is set to 0.

\paragraph{Balancing capacitor voltages} Various existing modulation techniques such as the nearest level modulation technique can be used. During operation, the capacitors can be charged or discharged depending on the direction of current flowing through the SMs. Consequently, the capacitor voltages can either increase or decrease independently causing unbalanced SM voltages when they are inserted. To mitigate this unbalance, the capacitor voltages need to be compared and regulated in a period of time. Denote the frequency of the regulation as $f_{\rm s}$, the time interval $\Delta t$ between any two voltage adjustments can be determined as $\Delta t = \frac{1}{f_{\rm s}}$. 

Various techniques for capacitor voltage regulation have already been proposed in the literature. In this paper, the sorting algorithm is used to balance the capacitor voltages. First, the SMs are sorted in ascending order within each arm according to the voltages of $v_{\rm c1}$. Thereafter, for arm $x$ the number of SMs needed to be inserted, denoted as $N_x$, will be calculated. Then, depending the direction of arm current $i_x$, which is equal to $i_k$, the operation mode of the SMs can be determined. When $i_x > 0$ and SM$_k$ is inserted, $v_{{\rm c1},k}$ increases, whereas $v_{{\rm c1},k}$ decreases when $i_x < 0$. Therefore, when $i_x > 0$, the first $N_x$ SMs in arm $x$ will be set in Mode 1 to be inserted, while the rest SMs are set in Mode 2 to be bypassed. On the contrary, the last $N_x$ SMs in the arm will be set in Mode 1 while the rest SMs are set in Mode 2 for $i_x < 0$. 

\paragraph{DC fault blocking} When MMC is overloaded or short-circuit faults occur during operation, the arm currents may excess the safety upper limit $I_{\rm F}$. To avoid damage and system failure, proper measurements need to be taken. As present in the previous section, the proposed SMs operating in Mode 3 have the ability to suppress short-circuit fault currents. Therefore, if $i_x > I_{\rm F}$, all SMs in arm $x$ will be set in Mode 3. After that, the capacitor voltages in these SMs increase and subsequently reduce the arm current. The SMs keep working in this mode until the load is alleviated or the fault is cleared.

\section{Experimental Verification} \label{experiment}
\begin{figure}[htb]
	\centering
	\includegraphics[width=6 cm]{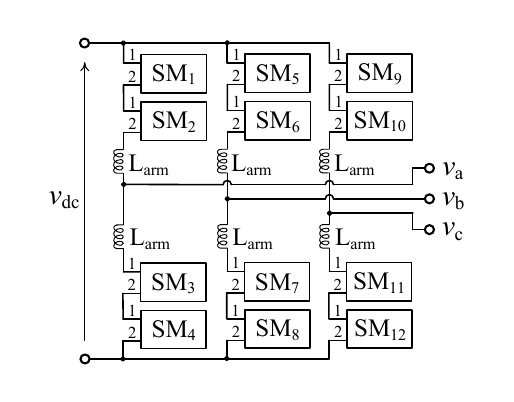}
	\caption{Circuit topology of the two-level MMC for the experiment.}
	\label{Experiment_Topology}
\end{figure}

\begin{figure}[htb]
	\centering
	\includegraphics[width=8.5 cm]{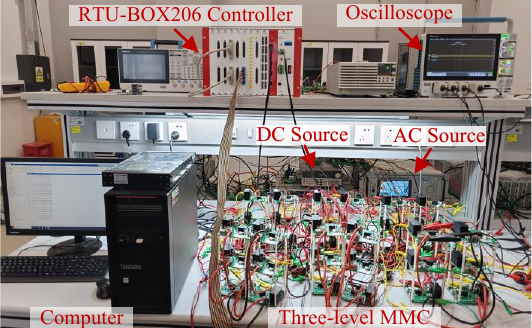}
	\caption{Practical setup of the experiment.}
	\label{Experiment_SetUp}
\end{figure}

The functionality of the proposed SM topologies at low voltage are verified by experiment where the SMs are applied to a two-level three-phase MMC as the circuit topology of the MMC is shown in Fig.~\ref{Experiment_Topology}, the experimental setup is shown in Fig.~\ref{Experiment_SetUp}, and the parameters used for the experiment are given in Table~\ref{Experiment_Parameters}. The nearest level control algorithm (NLC) \cite{Tu2011} is adopted and performed by an RTU-BOX206 controller from RTUnit to regulate the converter. Two scenarios are tested in the experiment: in the first scenario, the MMC operates normally as inverter or rectifier, respectively; in the second scenario, short-circuit fault occurs at the DC-side of the MMC.


\begin{table}[ht]
	\caption{Parameters used in the experiment.}
	\label{Experiment_Parameters} 
	\centering
	\vspace{-0.25 cm}
	\begin{tabular}{p{2cm}<{\centering} @{} p{1cm}<{\centering} | @{} p{3.5cm}<{\centering} @{} p{1cm}<{\centering}}
		\hline \hline
		Parameter & Value & Parameter & Value \\
		\hline
		$\rm C_1$ (${\rm \mu F}$) & 470 & $\rm C_2$ (${\rm \mu F}$) & 470 \\
		$\rm L_{arm}$ (mH) & 1 & $\rm R_{load}$ ($\Omega$) & 100 \\
		$v_{\rm dc}$ (V) & 10 & $v_{\rm ac}$ (ph-ph, RMS, V) & 8.66 \\
		\hline \hline
	\end{tabular}
\end{table}

\subsection{Scenario 1: normal operation}   
\begin{figure}[htb]
	\centering
	\includegraphics[width = 9 cm]{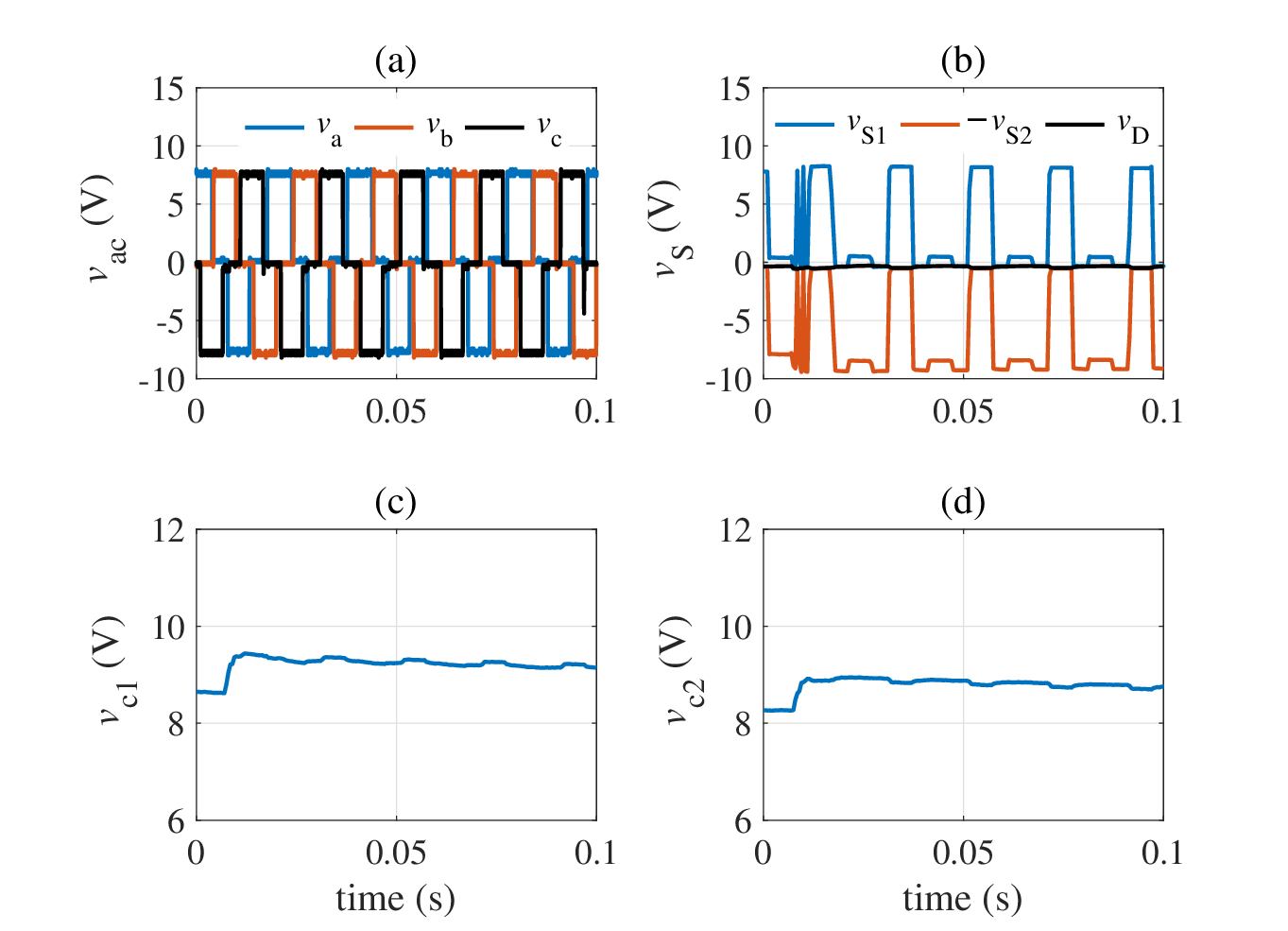}
	\caption{Experimental result of the MMC with the first SM variant operating in inverter mode: (a) AC output voltages; (b) voltages across the switches and diode in SM$_{12}$; (c) voltage across $\rm C_1$ in SM$_{12}$; (b) voltage across $\rm C_2$ in SM$_{12}$.}
	\label{Inverter_2S}
\end{figure}

\begin{figure}[htb]
	\centering
	\includegraphics[width = 9 cm]{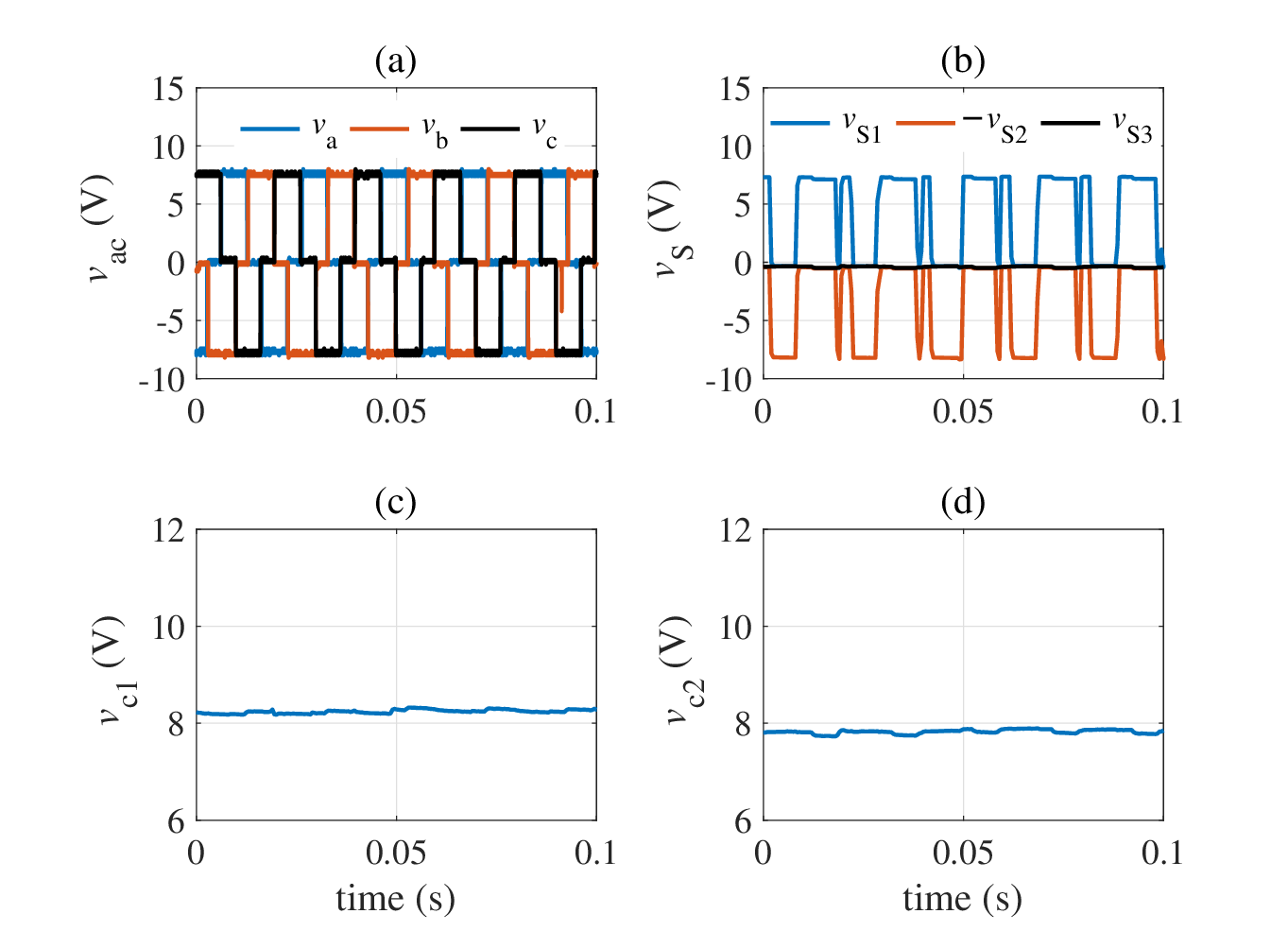}
	\caption{Experimental result of the MMC with the second SM operating in inverter mode: (a) AC output voltages; (b) voltages across the switches in  SM$_{12}$; (c) voltage across $\rm C_1$ in SM$_{12}$; (b) voltage across $\rm C_2$ in SM$_{12}$.}
	\label{Inverter_3S}
\end{figure}

In this scenario, the functionality of the MMC using the both proposed SM topologies operating in normal operation are tested. Two operation modes are considered. First, the MMC operates as an inverter, and the experiment results using SM variant 1 and variant 2 are measured and plotted in Fig. \ref{Inverter_2S} and Fig. \ref{Inverter_3S}, respectively. In this mode, its DC side is powered by a DC voltage source with output voltage of $v_{\mathrm{dc}}$, and each phase of the AC output is connected to a resistive load $\mathrm{R}_{\mathrm{load}}$. The AC output voltages of the MMC using SM variant 1 and variant 2 are plotted in Fig.~\ref{Inverter_2S}(a) and Fig.~\ref{Inverter_2S}(a), respectively. As can be seen from the figure, the MMC can generate symmetric three-phase three-level AC voltages using both SM topologies. During normal operation, switches S$_1$ and S$_2$ are successively turned-on and -off, while the diode in SM variant 1 and S$_3$ in SM variant 2 are kept conducting, as indicated by the voltages across the diode and the switches shown in Fig.~\ref{Inverter_2S}(b) and Fig.~\ref{Inverter_2S}(b). The capacitor voltages in both SMs are regulated around a constant value using the proposed balancing scheme, as shown in Fig.~\ref{Inverter_2S}(c), (d) and Fig.~\ref{Inverter_2S}(c), (d). 

\begin{figure}[htb]
	\centering
	\includegraphics[width = 9.5 cm]{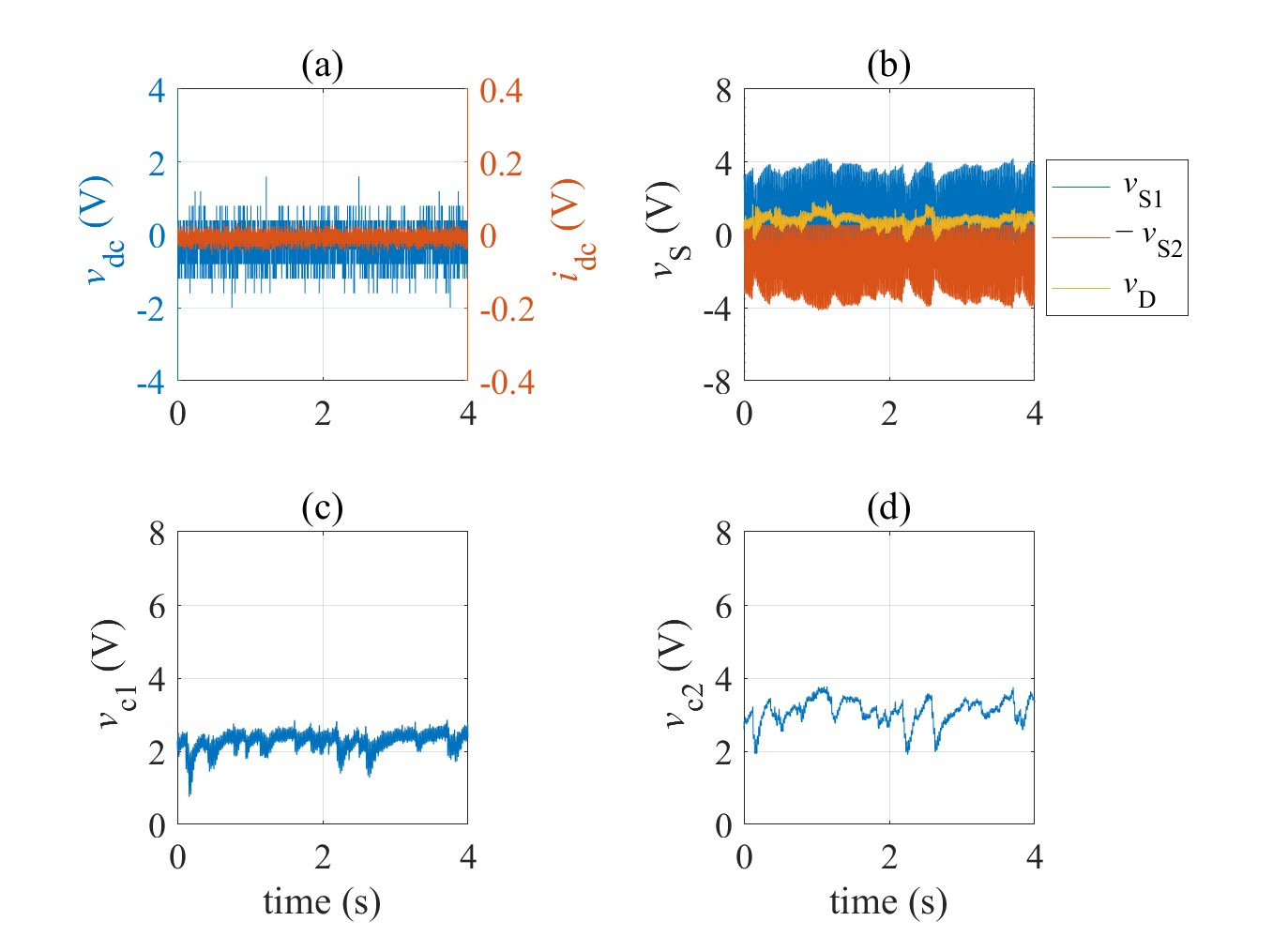}
	\caption{Experimental result of the MMC with the first SM operating in rectifier mode: (a) AC output voltages; (b) voltages across the switches and diode in SM$_{12}$; (c) voltage across $\rm C_1$ in SM$_{12}$; (b) voltage across $\rm C_2$ in SM$_{12}$.}
	\label{Rectifier_2S}
\end{figure}

\begin{figure}[htb]
	\centering
	\includegraphics[width = 9.5 cm]{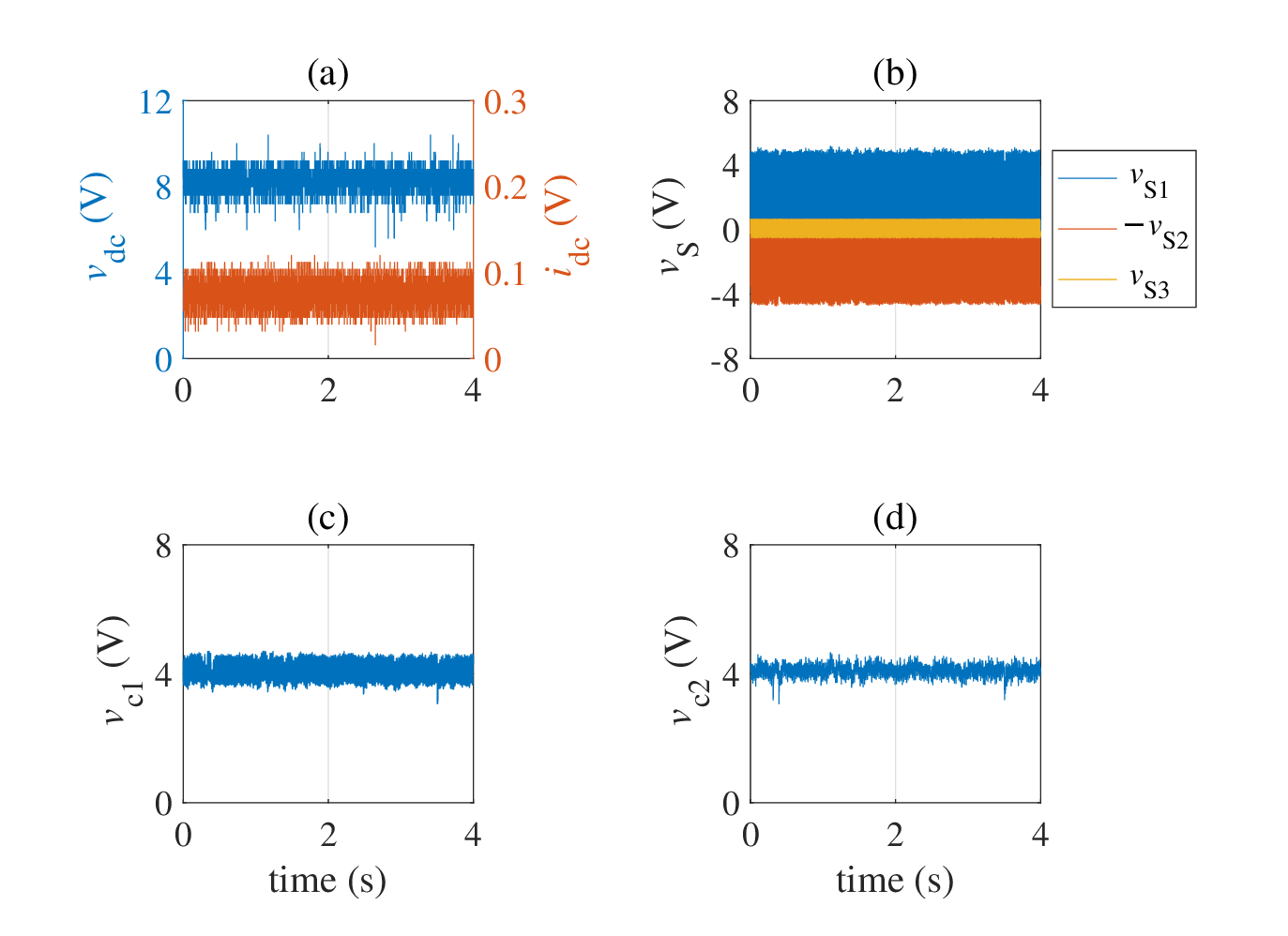}
	\caption{Experimental result of the MMC with the second SM operating in rectifier mode: (a) AC output voltages; (b) voltages across the switches in SM$_{12}$; (c) voltage across $\rm C_1$ in SM$_{12}$; (b) voltage across $\rm C_2$ in SM$_{12}$.}
	\label{Rectifier_3S}
\end{figure}

When the MMC operates as an rectifier, its AC side is powered by a three-phase AC voltage source with output voltage of $v_{\mathrm{ac}}$, and the DC output is connected to a resistive load $\mathrm{R}_{\mathrm{load}}$. The DC output voltage and current of the MMC using SM variant 1, are close to zero, as shown in Fig.~\ref{Rectifier_2S}(a). In contrast, the DC output voltage of the MMC based on SM variant 2 is about 8 V, as shown in Fig.~\ref{Rectifier_3S}(a). This indicates that only the SM variant 2 can be used in the rectification mode. Due to the unilateral conductivity of the diode in SM variant 1, current can not flowing from the AC-side to the DC-side of the MMC. Nevertheless, the voltages across all components in SM variant 2 remain within $\frac{1}{2}v_{\rm dc}$, as shown in Fig.~\ref{Rectifier_3S}(b)--(d). For SM variant 2, the capacitor voltages are kept around a constant value as shown in Fig.~\ref{Rectifier_3S}(c), (d), and the maximum voltages across the switches are close to the capacitor voltages. 

\subsection{Scenario 2: DC Fault}
\begin{figure}[htb]
	\centering
	\includegraphics[width=9 cm]{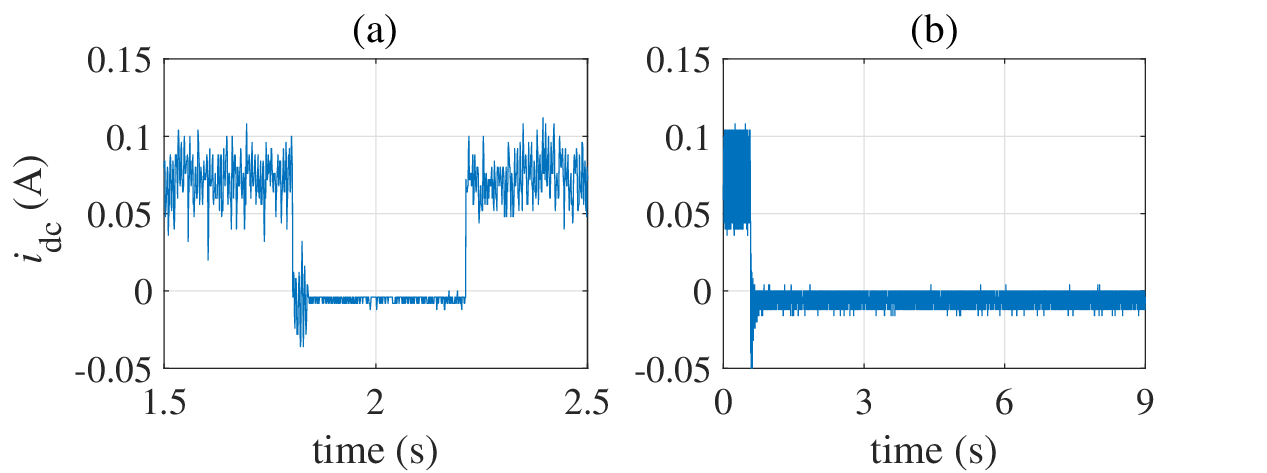}
	\caption{Response of DC current: (a) non-permanent DC fault; (b) permanent DC fault.}
	\label{I_dc}
\end{figure}

\begin{figure}[htb]
	\centering
	\includegraphics[width=9 cm]{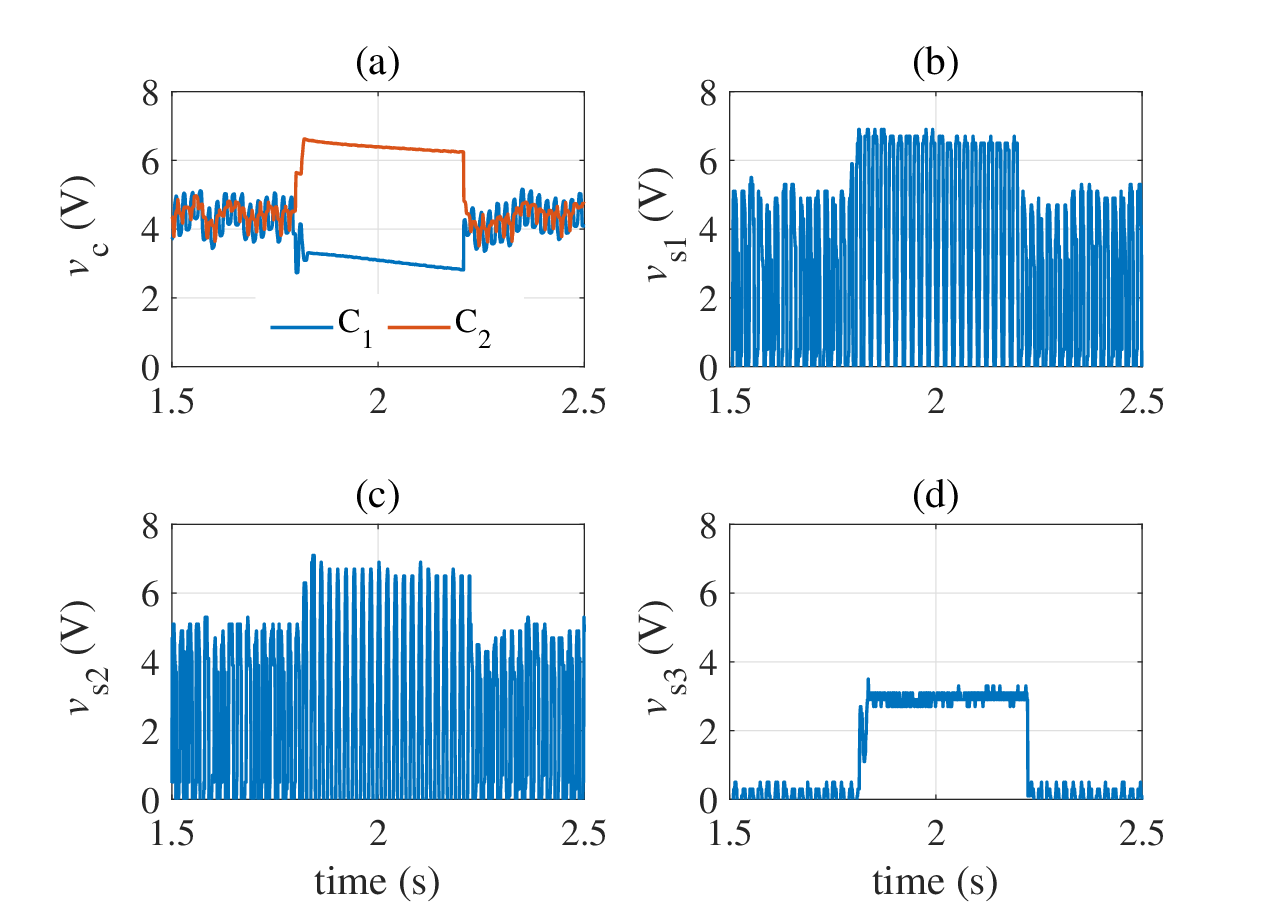}
	\caption{Voltage response of the capacitors and switches in all SMs of phase c during non-permanent DC fault.}
	\label{SM12}
\end{figure}

In this scenario, the fault blocking capability of the proposed SM is tested. Since SM variant 1 cannot operate in rectification mode, only the SM variant 2 is tested in the experiment, and the SM variant 1 is tested in the simulation in the next section. At around $t = 1.8$ s, a non-permanent short-circuit fault occurs at the DC-side of the MMC, all SMs in the MMC are immediately set to Mode 3 and the MMC operates in the blocked mode. Thereafter, the DC-voltage and -current quickly fall to zero, as shown in Fig. \ref{I_dc}(a), indicating that the fault current has been cleared. Meanwhile, the capacitor voltage of C$_1$ decreases, while the capacitor voltage of C$_1$ increases, as the capacitor voltages of SM$_{12}$ shown in Fig. \ref{SM12}(a). The voltages across the switches are within 1.5 times of their amplitude during normal operation, as shown in Fig. \ref{SM12}(b)--(d).

After the DC fault is blocked and the fault current has fallen to zeros, the SMs are reset back to Mode 1 or 2 at about $t = 2.2$ s, and the MMC quickly resumes normal operation. As shown in Fig. \ref{I_dc}(a), the DC current quickly increases back to the level as before the fault, and so as the switch voltages as shown in Fig. \ref{SM12}(b)--(d). 

\begin{figure}[htb]
	\centering
	\includegraphics[width=9 cm]{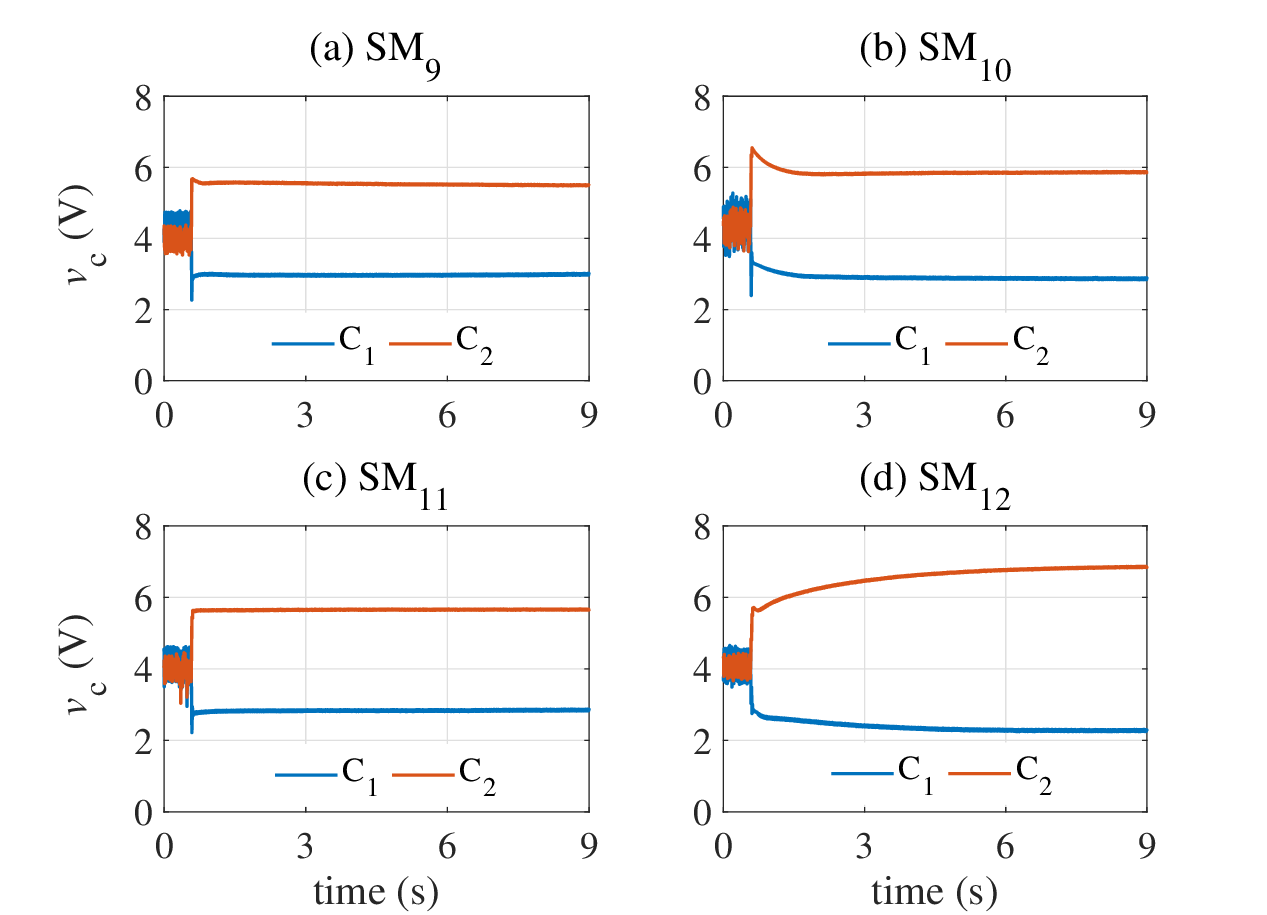}
	\caption{Voltage response of the capacitors in all SMS of phase c during permanent DC fault.}
	\label{Vc_PermanentFault}
\end{figure}

\begin{figure}[htb]
	\centering
	\includegraphics[width=9 cm]{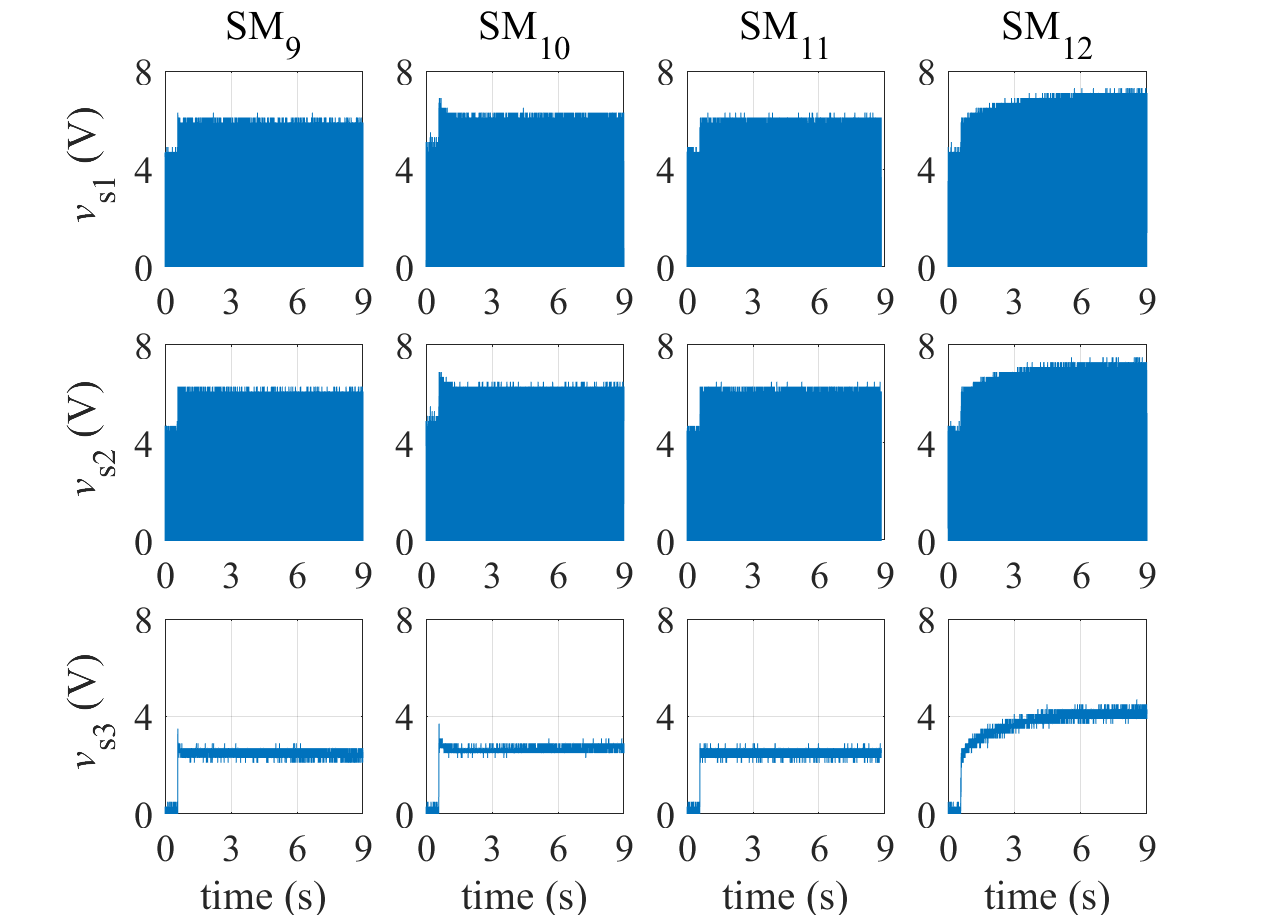}
	\caption{Voltage across the switches in all SMs of phase c during permanent DC fault.}
	\label{Vs_PermanentFault}
\end{figure}

To investigate if the voltage stress of each component will accumulate during DC fault, a further experiment of permanent DC fault is performed. Similar to the previous scenario, a short-circuit fault occurs at the DC-side of the MMC and the MMC is then blocked immediately at about $t = 0.57$ s. Consequently, the DC output current of the MMC quickly falls to zero as shown in Fig. \ref{I_dc}(b). Due to the system symmetry, only the voltage response of SM$_9$ -- SM$_{12}$ of phase C is visualized as shown in Fig. \ref{Vc_PermanentFault} and Fig. \ref{Vs_PermanentFault}. Similar to the previous scenario, the capacitor voltages of C$_1$ and C$_2$ decreases and increases in each SM, respectively, and become finally constant within two times of their amplitude in normal operation as can be seen from Fig. \ref{Vc_PermanentFault}. The maximum voltages across all switches are within two times of their maximum value during normal operation, as shown in Fig. \ref{Vs_PermanentFault}. 

\section{Application} \label{application}
\begin{figure*}[h!]
	\centering
	\includegraphics[width = 13 cm]{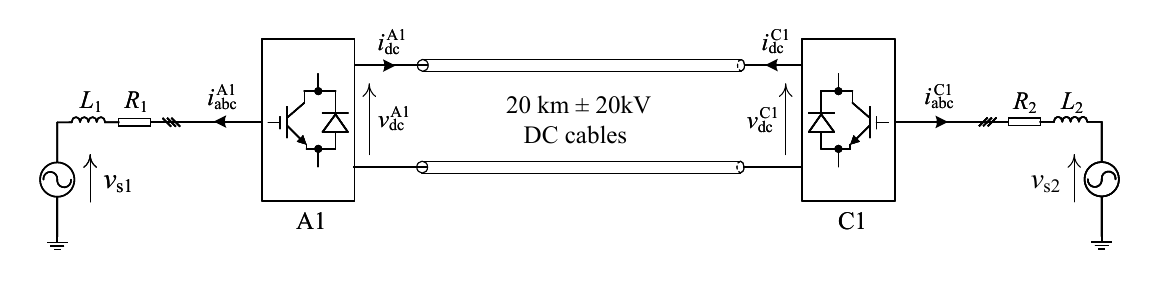}
	\caption{Topology of the HVDC-system used in the test simulation.}
	\label{Test_Simulation}
\end{figure*}

The proposed SM topologies were applied to an MMC-HVDC transmission system. The system refers to the VSC-HVDC system DCS1 in the CIGRE B4 DC Grid Test System \cite{Vrana2013}. The schematic of the system circuit topology is shown in Fig. \ref{Test_Simulation}. The DCS1 is a two-port symmetric monopole HVDC link with a voltage level of $\pm$200 kV. The DCS1 connects a 380-kV onshore AC system and a 145-kV offshore AC system through AC-DC converters and transformers. In this test simulation, each converter in the transmission system is implemented as a 101-level MMC converter. The proposed SM variant 1 in used in converter A1 operating as inverter, and the proposed SM variant 2 is adopted for converter C1 operating in rectification. The parameters used for the system are listed in Table \ref{Simulation_Parameters}. The test system is modeled and simulated using SimPowerSystem in MATLAB/Simulink 2021a.

\begin{table}[ht]
	\caption{Parameters used in the test simulation.}
	\label{Simulation_Parameters} 
	\centering
	\vspace{-0.25 cm}
	\begin{tabular}{p{3cm}<{\centering} @{} p{1cm}<{\centering} | @{} p{3cm}<{\centering} @{} p{1cm}<{\centering}}
		\hline \hline
		Parameter & Value & Parameter & Value \\
		\hline
		$\rm L_1$ (mH) & 3.977 & $\rm R_1$ ($\Omega$) & 0.0587 \\
		$\rm L_2$ (mH) & 6.522 & $\rm R_2$ ($\Omega$) & 0.0657 \\
		$\rm L_{arm}$ (mH) & 2.89 & $\rm R_{arm}$ ($\Omega$) & 0.5 \\
		$\rm C_{1,A1}$ (mF) & 20 & $\rm C_{2,A1}$ (mF) & 20 \\
		$\rm C_{1,C1}$ (mF) & 13 & $\rm C_{2,C1}$ (mF) & 13 \\
		$\rm L_{arm,A1}$ (mH) & 2.889 & $\rm R_{arm,A1}$ ($\Omega$) & 0.005 \\
		$\rm L_{arm,C1}$ (mH) & 2.889 & $\rm R_{arm,C1}$ ($\Omega$) & 0.005 \\
		$v_{\rm s1}$ (ph-ph, RMS, kV) & 22 & $v_{\rm s2}$ (ph-ph, RMS, kV) & 22 \\
		Rating $v_{\rm dc}$ (kV) & $\pm$20 & $N_{\rm SM}$ & 20 \\
		Rating power (MVA) & 80 & Cable length (km) & 20 \\
		$\rm L_{cable}$ (mH/km) & 2.112 & $\rm R_{cable}$ ($\Omega$/km) & 0.0095 \\
		$\rm C_{cable}$ ($\rm \mu$F/km) & 0.2104 & $\rm G_{cable}$ ($\rm \mu$S/km) & 0.0620 \\
		\hline \hline
	\end{tabular}
\end{table}

\begin{figure}[h!]
	\centering
	\includegraphics[width = 9 cm]{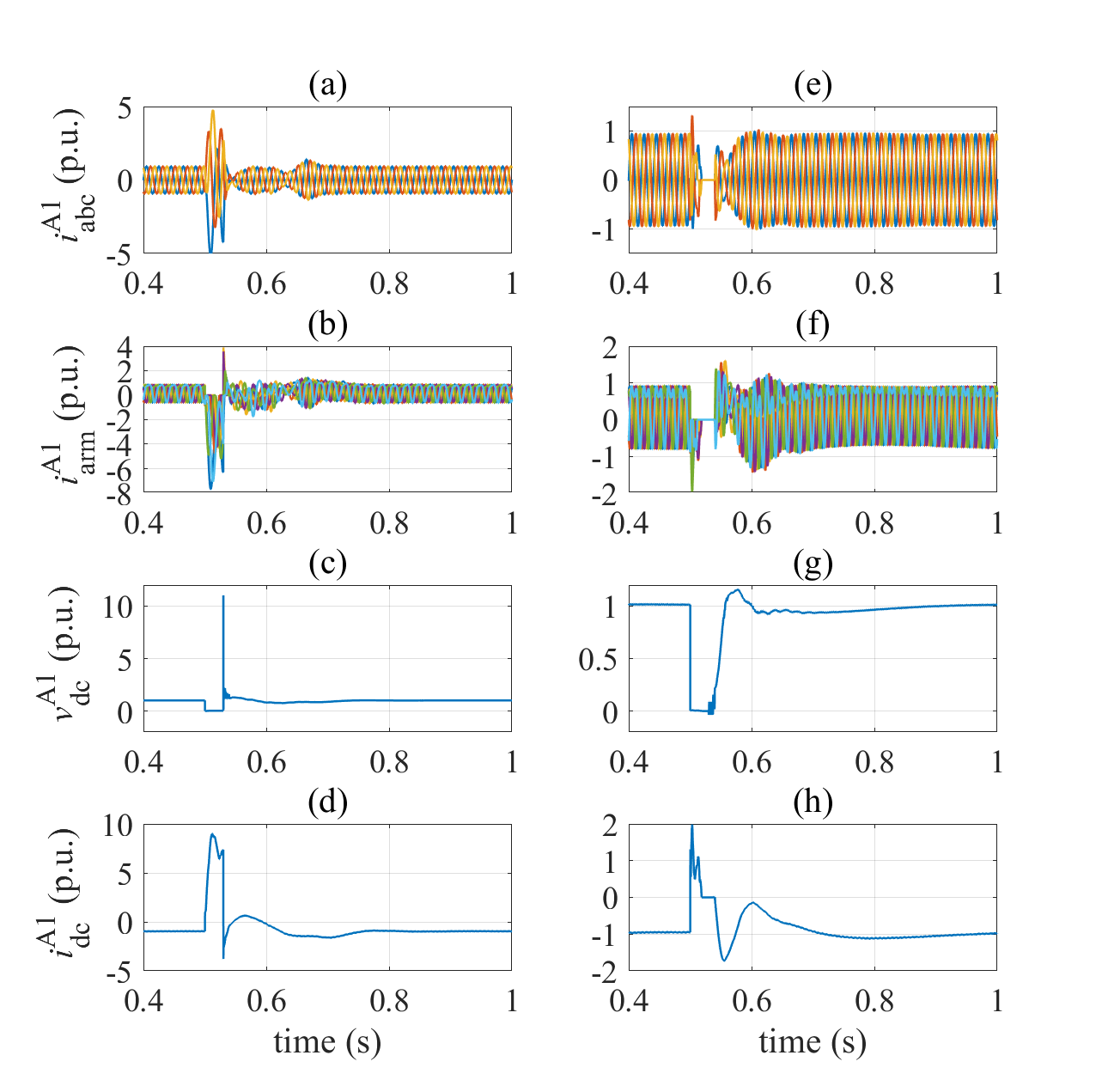}
	\caption{Comparison of response of MMC A1 using HBSM and the proposed SM variant 1 during non-permanent DC fault. HBSM: (a) AC-grid currents; (b) arm currents; (c) DC-link voltage; (d) DC-link current. The proposed SM: (e) AC-grid currents; (f) arm currents; (g) DC-link voltage; (h) DC-link current.}
	\label{VI_A1}
\end{figure}

\begin{figure}[h!]
	\centering
	\includegraphics[width = 9 cm]{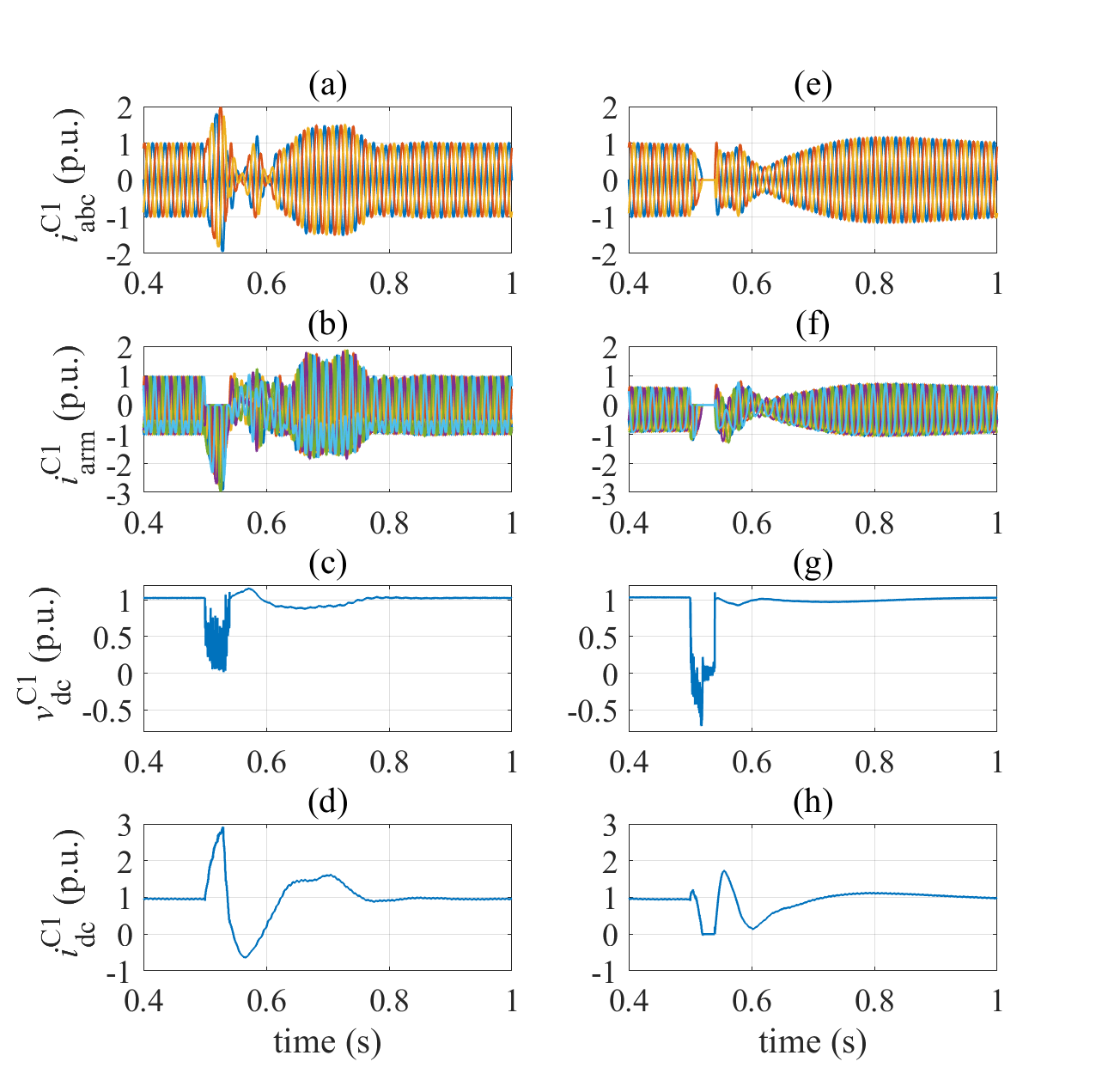}
	\caption{Comparison of response of MMC C1 using HBSM and the proposed SM variant 2 during non-permanent DC fault. HBSM: (a) AC-grid currents; (b) arm currents; (c) DC-link voltage; (d) DC-link current. The proposed SM: (e) AC-grid currents; (f) arm currents; (g) DC-link voltage; (h) DC-link current.}
	\label{VI_C1}
\end{figure}

After the system has reached steady state, a short-circuit fault occurs at the DC-side of A1 at $t = 0.5$ s. After that, the DC-link voltages at the DC-side of A1 and C1 fall quickly, as shown in Fig. \ref{VI_A1}(g) and Fig. \ref{VI_C1}(g). At the DC-side of both converters, the currents decrease quickly and increase reversely within 1 p.u. due to the discharge of the capacitors to the fault, as shown in Fig. \ref{VI_A1}(h) and Fig. \ref{VI_C1}(h). After the fault has been detected, both converters are set to blocked state by switching all SMs to Mode 3. Consequently, the DC-currents of both converters falls to zero, and so as the AC-currents of both converters as shown in Fig. \ref{VI_A1}(e) and Fig. \ref{VI_C1}(e), indicating that both proposed SMs have the capability to block DC fault currents. For comparison, the AC-currents and the DC-currents of both converters increase significantly using the HBSM, as shown in Fig. \ref{VI_A1}(a), Fig. \ref{VI_A1}(d), Fig. \ref{VI_C1}(a), and Fig. \ref{VI_C1}(d), respectively. 

\begin{figure}[h!]
	\centering
	\includegraphics[width = 9 cm]{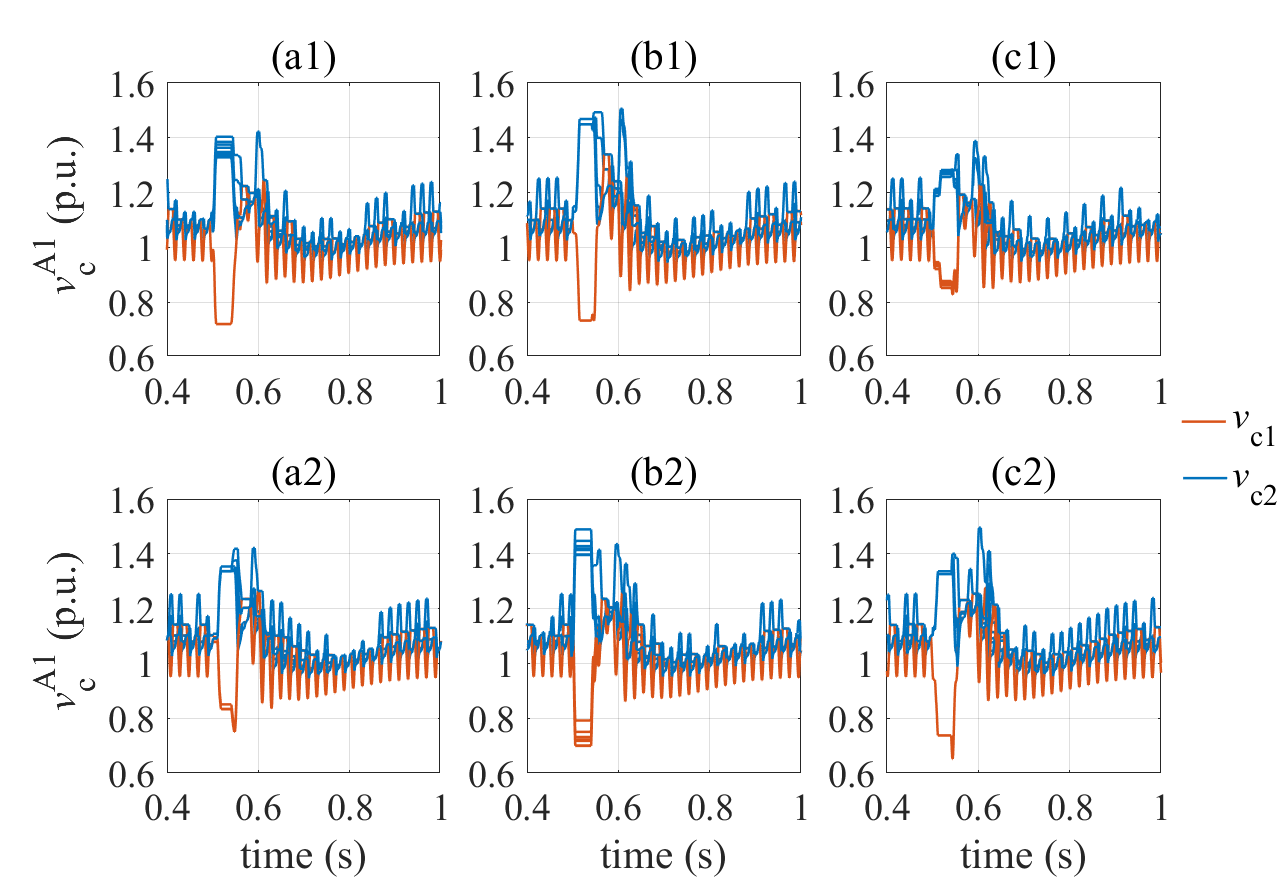}
	\caption{Voltage response of ALL capacitors in A1 during non-permanent DC fault in: (a1) upper arm of phase A; (a2) lower arm of phase A; (b1) upper arm of phase B; (b2) lower arm of phase B; (c1) upper arm of phase C; (c2) lower arm of phase C.}
	\label{Vc_A1_NonPermanentFault}
\end{figure}

\begin{figure}[h!]
	\centering
	\includegraphics[width = 9 cm]{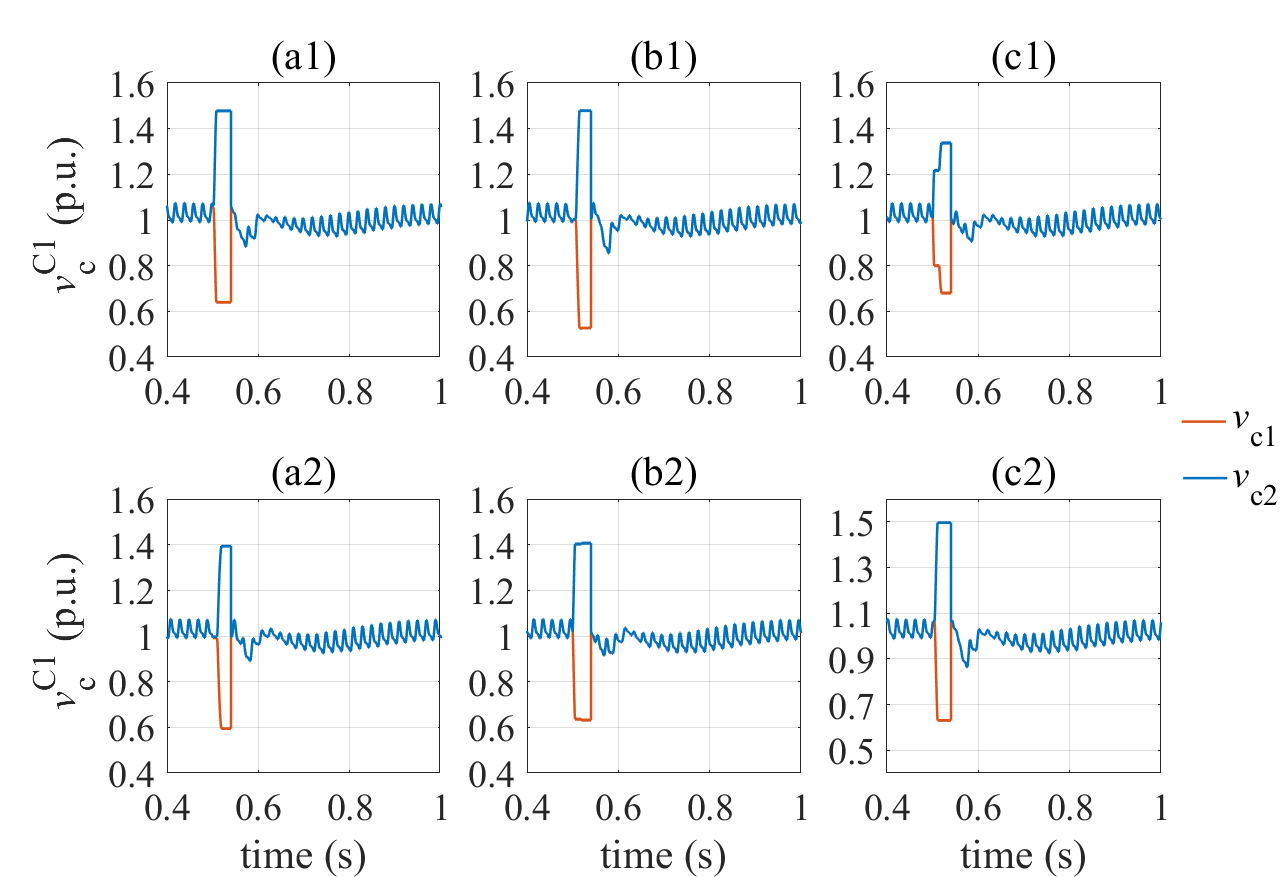}
	\caption{Voltage response of ALL capacitors in C1 during non-permanent DC fault in: (a1) upper arm of phase A; (a2) lower arm of phase A; (b1) upper arm of phase B; (b2) lower arm of phase B; (c1) upper arm of phase C; (c2) lower arm of phase C.}
	\label{Vc_C1_NonPermanentFault}
\end{figure}

\begin{figure}[h!]
	\centering
	\includegraphics[width = 9.5 cm]{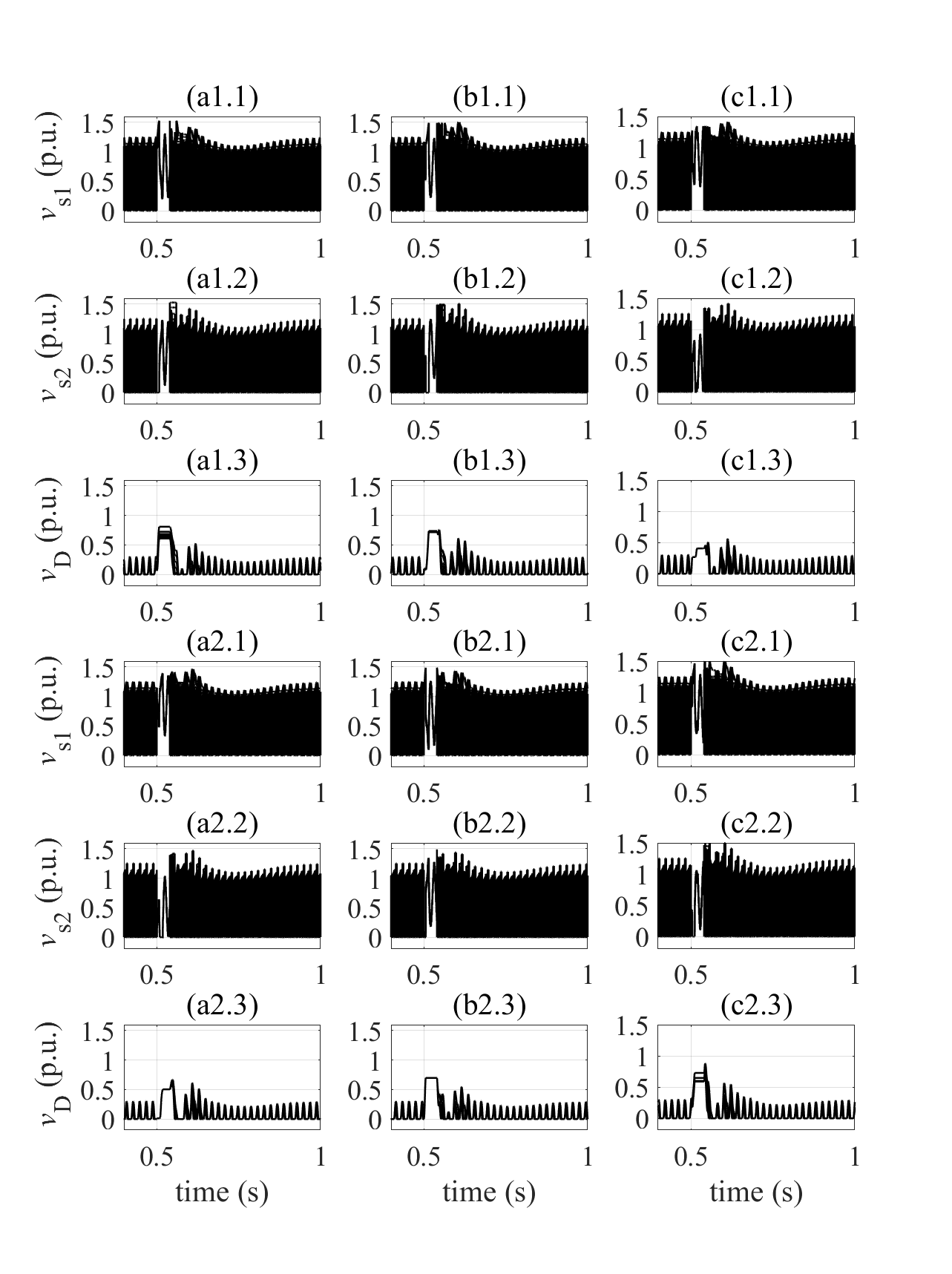}
	\caption{Voltages across the switches in A1 during non-permanent DC fault. Plot ($xi,j$) refers to all switches ${\rm S}_j$ in the upper arm ($i = 1$) or lower arm ($i = 2$) of phase $x$.}
	\label{Vs_A1_NonPermanentFault}
\end{figure}

\begin{figure}[h!]
	\centering
	\includegraphics[width = 9.5 cm]{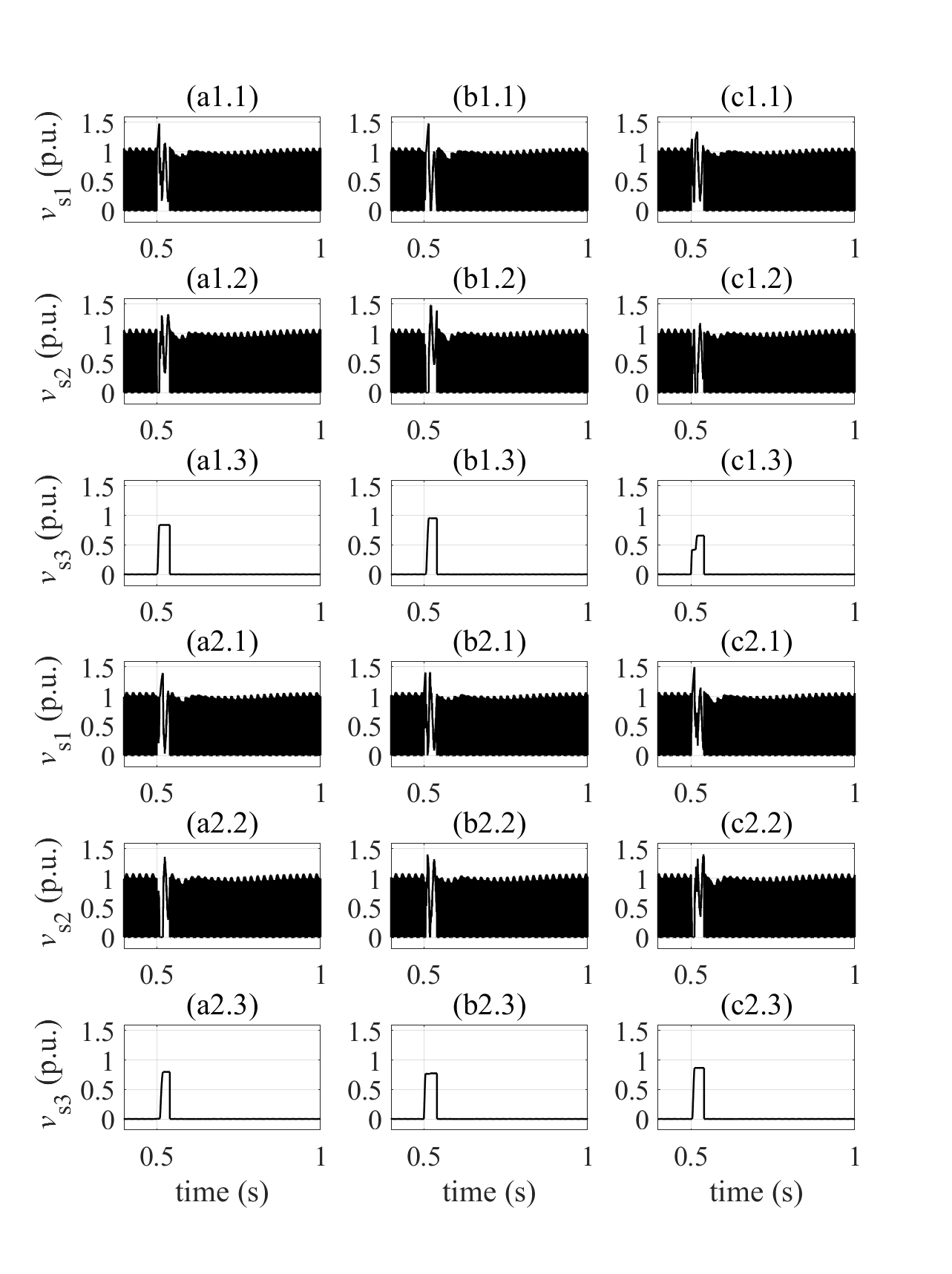}
	\caption{Voltages across the switches in C1 during non-permanent DC fault. Plot ($xi,j$) refers to all switches ${\rm S}_j$ in the upper arm ($i = 1$) or lower arm ($i = 2$) of phase $x$.}
	\label{Vs_C1_NonPermanentFault}
\end{figure}

During fault-blocking, the voltages of capacitors C$_1$ decrease whereas the voltages of capacitors C$_2$ increase for all SMs of both converters, as shown in Fig. \ref{Vc_A1_NonPermanentFault} and Fig. \ref{Vc_C1_NonPermanentFault}. After reaching the maximum value within 1.5 p.u. in general, the capacitor voltages stay in constant as the arm currents become zero in both converter, as shown in Fig. \ref{VI_A1}(b) and Fig. \ref{VI_C1}(b). The voltages across S$_1$ and S$_2$ are within 1.5 p.u., as shown in Fig. \ref{Vs_A1_NonPermanentFault} and Fig. \ref{Vs_C1_NonPermanentFault}. Since the diodes in converter A1 are in reversed bias, and the switches S$_3$ in converter C1 are turned off, therefore, the voltages across them are equal to the difference of both capacitor voltages in each SM, as shown in Fig. \ref{Vs_A1_NonPermanentFault} and Fig. \ref{Vs_C1_NonPermanentFault}. As can be seen from the figures, all switch voltages are within 1.5 p.u., indicating an acceptable voltage stress for all switches during the fault-blocking.  

At about $t = 0.52$ s, the fault currents are extinguished, the short-circuit fault has been removed, and both converters are reset to normal operation at $t = 0.54$ s. Thereafter, the DC-link voltages and the DC-currents of both converters are gradually rebuilt to the level before the fault occurred, as shown in Fig. \ref{VI_A1}(g)--(h) and Fig. \ref{VI_C1}(g)--(h). Meanwhile, the capacitor voltages and the voltages across all switches in all SMs return to the level as before the fault occurrence during the recovery, as shown in Fig. \ref{Vc_A1_NonPermanentFault}, Fig. \ref{Vc_C1_NonPermanentFault}, Fig. \ref{Vs_A1_NonPermanentFault}, and Fig. \ref{Vs_C1_NonPermanentFault}.

\section{Conclusion} \label{conclusion}
A topology of SM with two variants and control scheme for MMC were proposed in this paper. Both variants have the capability to block DC-fault current without auxiliary protective equipment such as circuit breaker. The functionality of the proposed SMs was verified by physical experiment, and their value has been demonstrated by simulation. The exiting topologies with the same capability require at least three fully-controlled switches in each SM. One proposed SM topology contains three fully-controlled switches, the second SM contains only two switches as the traditional HBSM but featured with fault-blocking capability which the HBSM lacks. Compared to a DCSM with the same component, both SMs can offer greater energy storage capacity. These merits can help reduce the building cost and operation complexity for MMCs.

\bibliography{IEEEabrv,References}

\vfill

\end{document}